\documentclass[10pt,journal,compsoc]{IEEEtran}

\def\BibTeX{{\rm B\kern-.05em{\sc i\kern-.025em b}\kern-.08emT\kern-.1667em\lower.7ex\hbox{E}\kern-.125emX}}

\usepackage[space]{cite}
\usepackage{amsmath,amssymb,amsfonts}
\usepackage{algorithmic}
\usepackage{graphicx}
\usepackage{textcomp}
\usepackage{xcolor}
\usepackage{color,colortbl}
\definecolor{Gray}{gray}{0.9}
\usepackage{graphicx,tabu,tabularx}
\usepackage{xspace}
\usepackage{bm}
\usepackage{color}
\usepackage{breakurl}
\PassOptionsToPackage{hyphens}{url}\usepackage{hyperref}
\usepackage[font=small]{subcaption}

\usepackage{multirow}
\usepackage{enumitem}
\usepackage{booktabs}

\usepackage{balance}
\begin{document}

\graphicspath{ {images/}}
\newcommand{\sysname}{CNN\-Select\xspace}
\newcommand{\modiselect}{CNN\-Select\xspace}

\newcommand{\para}[1]{\noindent \textbf{#1}}

\title{Characterizing the Deep Neural Networks Inference Performance of Mobile Applications}

\author{Samuel S. Ogden, Tian Guo%
\IEEEcompsocitemizethanks{\IEEEcompsocthanksitem
Ogden and Guo are with the department of Computer Science, Worcester Polytechnic Institute, MA, 01609.\protect\\
Email: ssogden@wpi.edu, tian@wpi.edu
}% <-this % stops an unwanted space
\thanks{This work is supported in part by National Science Foundation grants \#1755659 and \#1815619 and Google Cloud Platform Research credits. Manuscript received September 9, 2019; revised month day, year.}}

\IEEEtitleabstractindextext{
\begin{abstract}
Today's mobile applications are increasingly leveraging deep neural networks
to provide novel features, such as image and speech recognitions. To use a
pre-trained deep neural network, mobile developers can either host it in a cloud
server, referred to as \emph{cloud-based inference}, or ship it with their mobile
application, referred to as \emph{on-device inference}. In this work, we
investigate the inference performance of these two common approaches on both mobile devices and public clouds, using popular
convolutional neural networks.
Our measurement study suggests the need for both on-device and
cloud-based inferences for supporting mobile applications. In particular,
newer mobile devices is able to run mobile-optimized CNN models in
reasonable time. However, for older mobile devices or to use more complex
CNN models, mobile applications should opt in for cloud-based inference. We
further demonstrate that variable network conditions can lead to poor cloud-based
inference end-to-end time. To support efficient cloud-based inference, we propose
a CNN model selection algorithm called \modiselect that dynamically selects the
most appropriate CNN model for each inference request, and adapts its
selection to match different SLAs and execution time budgets that are caused
by variable mobile environments. The key idea of \modiselect is to make
inference speed and accuracy trade-offs at runtime using a set of CNN
models. 
We demonstrated that \modiselect smoothly improves inference accuracy while maintaining SLA attainment in 88.5\% more cases than a greedy baseline.
\end{abstract}

}

\maketitle

\section{Introduction}
\label{sec:intro}

Resource intensive deep learning models are increasingly used in mobile applications~\cite{NIPS2013_5004, DBLP:journals/corr/WuSCLNMKCGMKSJL16} to add features such as real-time language translation, image recognition and personal assistants~\cite{NIPS2013_5004, DBLP:journals/corr/WuSCLNMKCGMKSJL16, siri.deeplearning}. To use deep learning models, mobile applications can
either utilize cloud-based inference services or run directly on-device.
However, achieving faster deep inference with high accuracy is often constrained
by mobile computation, storage and network conditions.

In this work, we ask the question: what are the performance trade-offs of using on-device
versus cloud-based inference for mobile applications. We conduct an empirical
measurement study to quantify both the end-to-end inference time and resource
consumption under different setups. In particular, we identify a number of key
factors, including deep learning frameworks, mobile devices, and CNN model
compression techniques, that impact on-device inference.
Further, we study the cloud-based inference performance under different mobile
network conditions and using different cloud servers. We identify that model
startup latency can impose orders of magnitude time overhead and should
be properly considered when managing cloud inference servers.

In sum, our observations reveal the relatively large, though shrinking, performance gaps between on-device and
cloud-based inference. As such, it is still preferable for older mobile devices that require
more complex CNN models to resort to cloud-based inference. To support
this efficient cloud-based inference under dynamic mobile network
conditions, we propose \modiselect to manage and
select the most appropriate CNN model for each mobile inference request.
The key insight for
designing \modiselect is that each model in a set of CNN models exposes
different inference time and accuracy trade-offs that can be leveraged to mask the inherent
mobile network variability.
The dynamic model selection algorithm frees mobile
developers from specifying CNN models before deployment and the
probabilistic-nature makes it more robust towards unpredictable performance
variations such as increased inference time due to sudden workload spikes.

We make the following contributions.
\begin{itemize}[leftmargin=*]
\item \textbf{Performance characterization of mobile deep inference.} We conducted
an extensive empirical performance analysis of using CNN models in mobile applications. We
identified key performance factors, e.g., mobile hardware and network conditions, and quantified their impacts for both on-device and cloud-based inference.
\item \textbf{Performance-aware CNN models selection algorithm.} We propose \modiselect, a probabilistic-based algorithm that
chooses the best CNN model to account for dynamic mobile environments when using
cloud-based inference. We conducted an end-to-end evaluation and extensive
empirical-driven simulations that demonstrate \modiselect's ability to smoothly
trade-off between inference accuracy and time.
\end{itemize}
\section{Background}
\label{sec:bg}

\subsection{Deep Learning Models}

\begin{figure}[t]
\centering
\includegraphics[width=.4\textwidth]{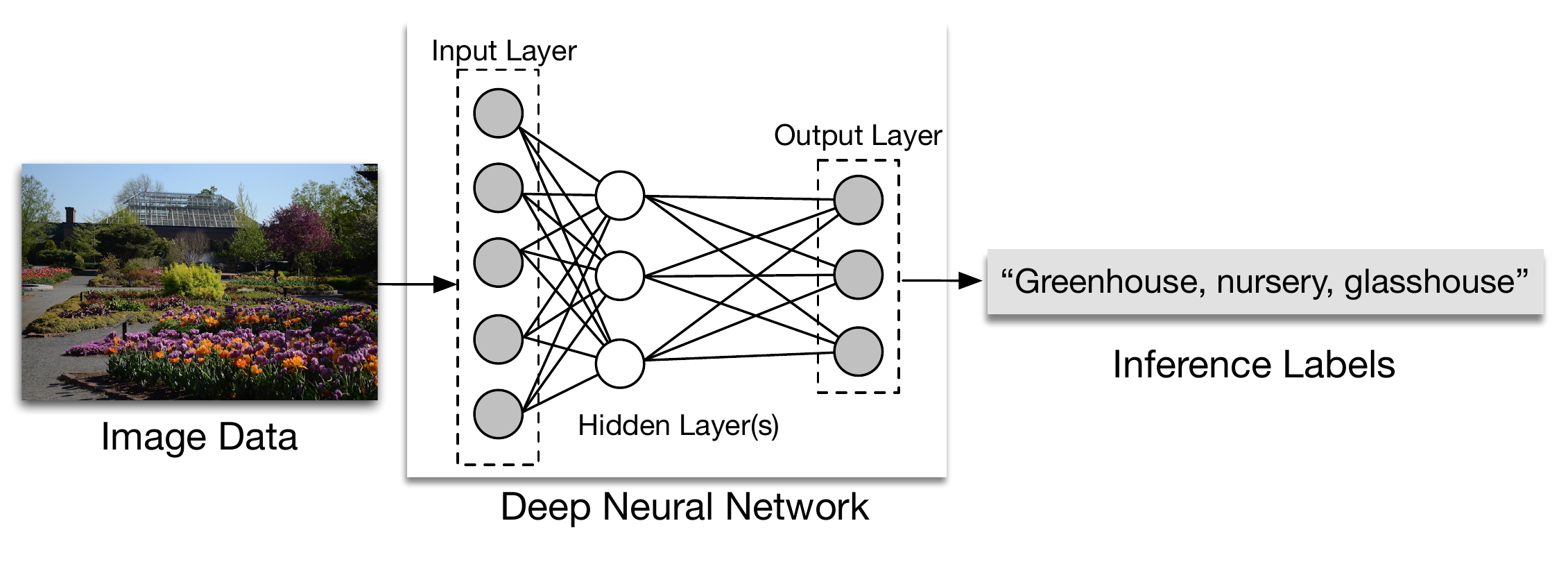}
\caption{\textbf{Image recognition with deep neural networks.}
An image passes through a deep neural network that consists of layers of neurons. The output layer produces the inference labels that best describe the image.}
\label{fig:object_recognition}
\end{figure}

Deep learning refers to a class of artificial neural networks (ANNs) for
learning representations of input data utilizing many layers and are widely used for visual
and speech recognition~\cite{NIPS2012_4824, nasnet, 6638947, Goodfellow-et-al-2016}.
In this paper, we focus on a special class of deep learning models called convolutional neural networks (CNNs)~\cite{NIPS2012_4824,nin:2014,SqueezeNet:2016}.
CNNs are widely used for visual recognition tasks, e.g., in Figure~\ref{fig:object_recognition}, and have been seen to exhibit high accuracy in such tasks.
CNNs usually consist of a number of convolution layers followed by other layer types, such as pooling layers and fully connected layers.
Each layer takes data from previous layer, processed through non-linear activation functions, and apply predefined computation. % in parallel.
More specifically, convolutional layers generate feature maps~\cite{lecun-01a}, while pooling layers control overfitting by reducing parameters in the representation~\cite{ws:cs231n}.

There are a number of popular deep learning frameworks~\cite{caffe,caffe2,collobert2002torch,tensorflow2015-whitepaper} that ease the training and deploying of deep learning models.
Different frameworks require different syntaxes to describe CNNs and have different trade-offs for training and inference~\cite{bahrampour2015comparative} phases.
In this paper, we look at pre-trained CNN models~\cite{ws:caffeModelZoo} supported by two popular deep learning frameworks: Caffe and TensorFlow.
A pre-trained model often consists of a binary model file and a text file that describe the model parameters and the network respectively, as well as text files of the output labels.

\subsection{Mobile Deep Inference}

\begin{figure}[t]
\centering
\begin{subfigure}{.36\textwidth}
\includegraphics[width=\columnwidth]{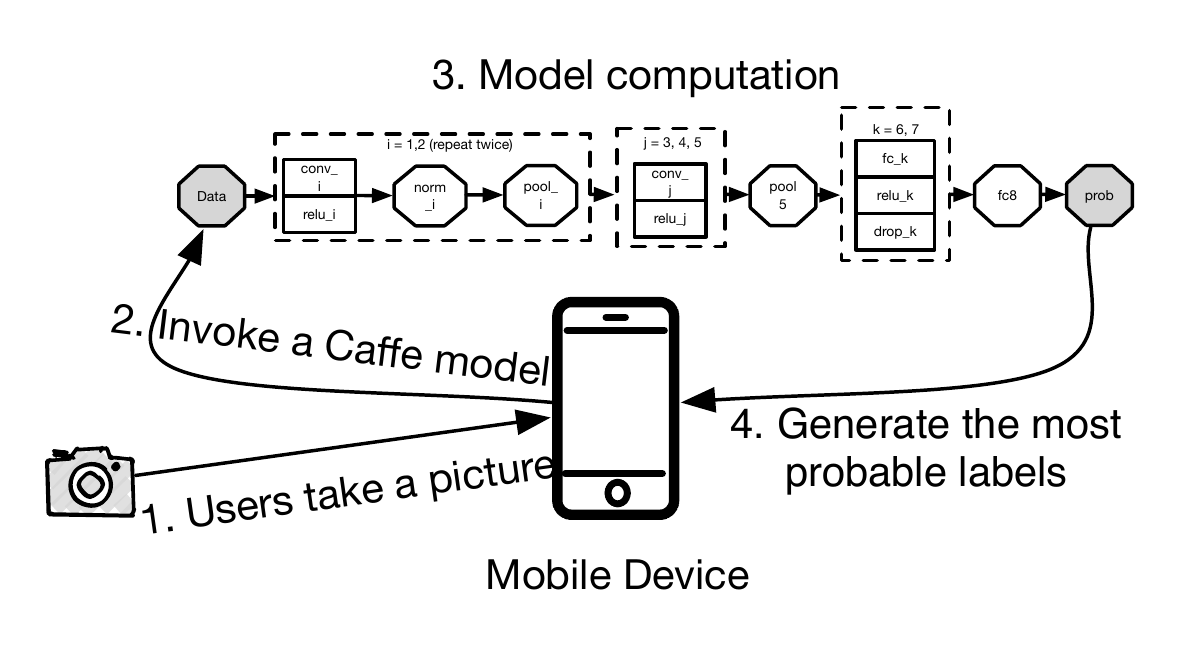}
\caption{\textnormal{On-device inference.}}
\label{subfig:ondevice_inference}
\end{subfigure}
\hfill
\begin{subfigure}{0.4\textwidth}
\includegraphics[width=\columnwidth]{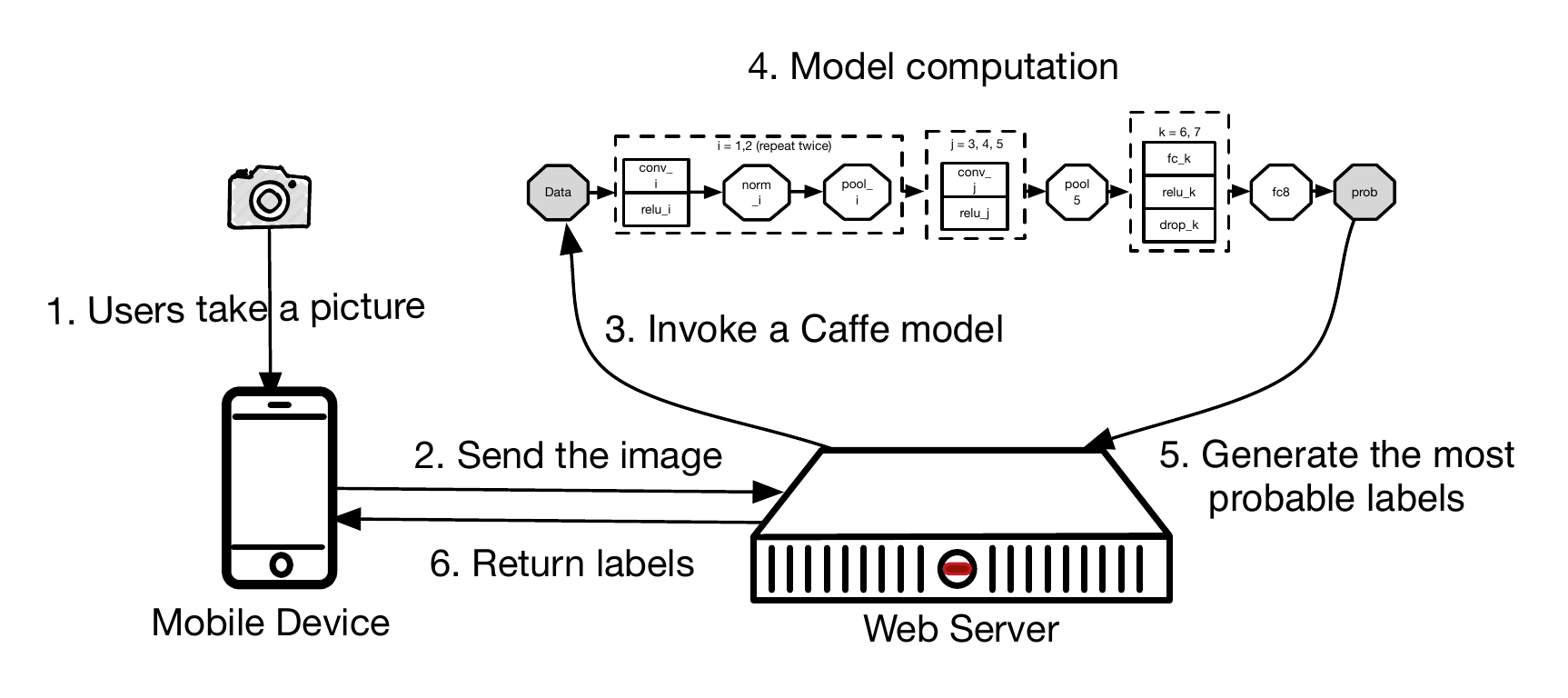}
\caption{Cloud-based inference.}
\label{subfig:cloud_inference}
\end{subfigure}
\caption{
\textbf{Design choices of deep learning powered mobile applications.} We use our implemented object recognition Android App as an example to illustrate the steps involved to perform on-device and cloud-based inference.}
\label{fig:mobile_deep_inference}
\end{figure}

\begin{figure}[t]
\centering
\includegraphics[width=.4\textwidth]{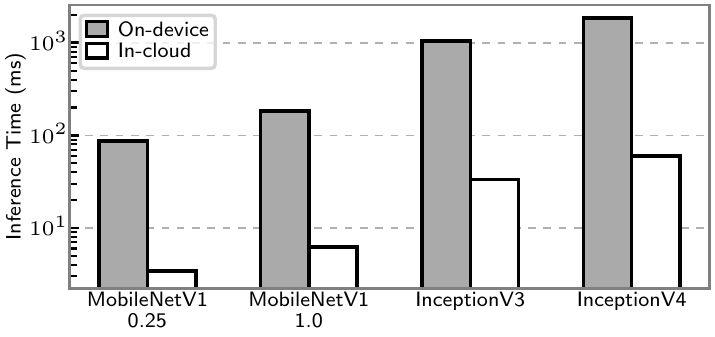}
\caption{
\textbf{Inference time of on-device and cloud-based inference.}
We measured the inference time of four CNN models running on a
high-end mobile device and an AWS \texttt{p2.xlarge} server.
}
\label{fig:model_latency}
\end{figure}

Mobile deep inference is defined as mobile applications using deep learning models to provide novel application features, such as image recognition.
Mobile deep inference can happen either directly \emph{on-device} or on \emph{cloud-based} servers~\cite{2018ic2e:guo}, as illustrated in Figure~\ref{fig:mobile_deep_inference}.

\para{On-device inference.}
A number of deep learning frameworks, such as Caffe2~\cite{caffe2} and
TensorFlow Lite~\cite{google_tensorflow:lite}, support executing deep learning
models directly on mobile devices. These frameworks
perform inference using exported models that have trained on powerful servers.
Even with the optimizations in these software libraries, on-device inference can
still be orders of magnitude slower than running inference on powerful servers
(shown in Figure~\ref{fig:model_latency} ).
These large performance gaps are mainly due to constraints on mobile hardware, e.g., lacking GPUs and having relatively little memory.
The inference inefficiency is exacerbated when an application needs a more
accurate model or needs to load multiple models, e.g., chaining the execution of
an optical character recognition model and a text translation
model~\cite{google_translate:ocr, google_translate:one_shot}.
To ensure mobile inference execution completes in a reasonable time, mobile-specific models~\cite{DBLP:journals/corr/HowardZCKWWAA17, nasnet} often sacrifice inference accuracy and thus may exclude complex application scenarios.
Even though on-device inference is a plausible choice for simple tasks and newer mobile devices, it is less suitable for complex tasks and older mobile devices.

\para{Cloud-based inference.}
Instead of on-device execution, mobile applications can send their inference requests to cloud-hosted models.
A number of model serving systems~\cite{clipper, tensorflow_serving, Gao:2018:LLR:3190508.3190541} have been proposed to manage different model versions available for executing inferences.
These systems often focus on maximizing inference throughput by batching incoming requests.
This may increase waiting time of some requests and consequently have negative impacts on end-to-end inference requests.

To utilize such systems, developers need to \emph{manually} specify the exact DNN model to use through exposed API endpoints.
For mobile developers, this manual model selection fails to consider the impact
of dynamic mobile network conditions, which can take up a significant portion of
end-to-end inference time~\cite{MODI, Satyanarayanan:2009:CVC:1638591.1638731}.
Such static development-time efforts can lead to picking a faster but less
accurate DNN model or risking SLA violations by choosing a more sophisticated DNN model.

Cloud-based inference has the potential to support a plethora of application scenarios, simple and complex, and heterogeneous mobile devices, old and new.
However, current mobile-agnostic serving platforms fall short in automatically adapting inference accuracy to varying time requirements of mobile inference requests.

\section{Understanding on-device Inference}
\label{sec:measure_ondevice}

In this section, we perform an empirical measurement of on-device inference time
and study the impact of key factors such as deep learning frameworks, mobile
capacities, CNN models and model compression techniques. We then analyze the
resource and energy implications of running on-device inference. 

\begin{table}[t]
\centering
\resizebox{0.48\textwidth}{!}{%
\begin{tabular}{@{}r|rrrrrr@{}}
\toprule
\textbf{\begin{tabular}[c]{@{}r@{}}mobile \\ device\end{tabular}} & \textbf{OS version} & \textbf{CPU} & \textbf{GPU} & \textbf{\begin{tabular}[c]{@{}r@{}}Memory\\ (GB)\end{tabular}} & \textbf{\begin{tabular}[c]{@{}r@{}}Storage\\ (GB)\end{tabular}} & \textbf{\begin{tabular}[c]{@{}r@{}}Battery\\ (mAh)\end{tabular}} \\ \midrule
\textbf{Nexus 5} & Android 6.0  & \begin{tabular}[c]{@{}r@{}}2.26 GHz\\ quad-core\end{tabular}      & \begin{tabular}[c]{@{}r@{}}129.8 GFLOPs\\ Adreno 330\end{tabular} & 2 & 16 & 2300 \\
\textbf{LG G3} & Android 6.0    & \begin{tabular}[c]{@{}r@{}}2.5 GHz\\ quad-core\end{tabular}       & \begin{tabular}[c]{@{}r@{}}129.8 GFLOPs\\ Adreno 330\end{tabular} & 3 & 32 & 3000  \\
\textbf{Moto G5 Plus} & Android 7.0  & \begin{tabular}[c]{@{}r@{}}2.0Gz\\ octa-core\end{tabular}         & \begin{tabular}[c]{@{}r@{}}130 GFLOPs\\ Adreno 506\end{tabular}   & 3 & 32 & 3000 \\
\textbf{Pixel 2} & Android 9.0  & \begin{tabular}[c]{@{}r@{}}4x 2.35GHz\\ 4x 1.9GHz\end{tabular}     & \begin{tabular}[c]{@{}r@{}}567 GFLOPs\\ Adreno 540\end{tabular}   & 4 & 128 & 2700 \\ \bottomrule
\end{tabular}%
}
\caption{\textbf{Mobile devices used in on-device measurement.}  We used four mobile devices that run their respectively most up-to-date OSes and have variable hardware resource capacities. Among which, Pixel 2 has a much faster mobile GPU.}
\label{tab:on_device_measurement_setup}
\end{table}

\para{Measurement methodology.}
We implemented two Android-based mobile image classification applications
and evaluated the on-device inference performance with four mobile devices
outlined in Table~\ref{tab:on_device_measurement_setup}. The first application
compared two mobile inference execution frameworks, Caffe2~\cite{caffe2} and
CNNDroid~\cite{cnndroid:2016} in order to compare setup and execution time of
CNNs on the Nexus5 device. The second application leveraged TensorFlow Mobile~\cite{google_tensorflow:mobile} in order to examine the impact of various model architectures and optimizations across all of our mobile devices.

For these tests we used a variety of CNN models.
For the first application, we used AlexNet~\cite{NIPS2012_4824}, NIN~\cite{nin:2014}, and SqueezeNet~\cite{SqueezeNet:2016}.
These three models were chosen because they have similar top-5 accuracy on the ImageNet dataset~\cite{imagenet:accuracy} but differ vastly in terms of model size and complexity.
The second application used four models that consisted of two accuracy-optimized
models, i.e., \texttt{InceptionV3} and
\texttt{InceptionV4}~\cite{DBLP:journals/corr/SzegedyVISW15}, and two time-optimized
models, i.e., \texttt{MobileNetV1 0.25} and \texttt{MobileNetV1 1.}~\cite{ws:mobilenets}.
These models were chosen for exposing a range of accuracy and complexity
trade-offs. Using the second application we also did an in-depth study of the
InceptionV3 model and optimizations thereof.

We furthermore used two different image datasets for generating inference requests.
For the first application we used an image set (\emph{images\_1}) that consists of 15 images.
For the second application we used a training set of 3314 to retrain our target
models and another 1000 images (\emph{images\_2}) to evaluate inference accuracy
and time.

For each experiment we break the end-to-end inference time into three parts: loading input image, loading CNN model, and CNN model execution time.
In the first application we output these time intervals to the Android log file which we accessed with \texttt{Logcat}, a command line tool, allowing us to additionally collect Android Runtime (ART) garbage collection information and application-level logs.
We additionally measured the power and resource consumption with the Trepn
profiler~\cite{ws:trepn} to sample battery level, normalized CPU and GPU load
every 100ms, following best practice techniques to minimize profile impact.
Our second application recorded time intervals to an SQLite database.

\begin{figure*}[t]
\centering
\begin{subfigure}{0.28\textwidth}
\includegraphics[width=\columnwidth]{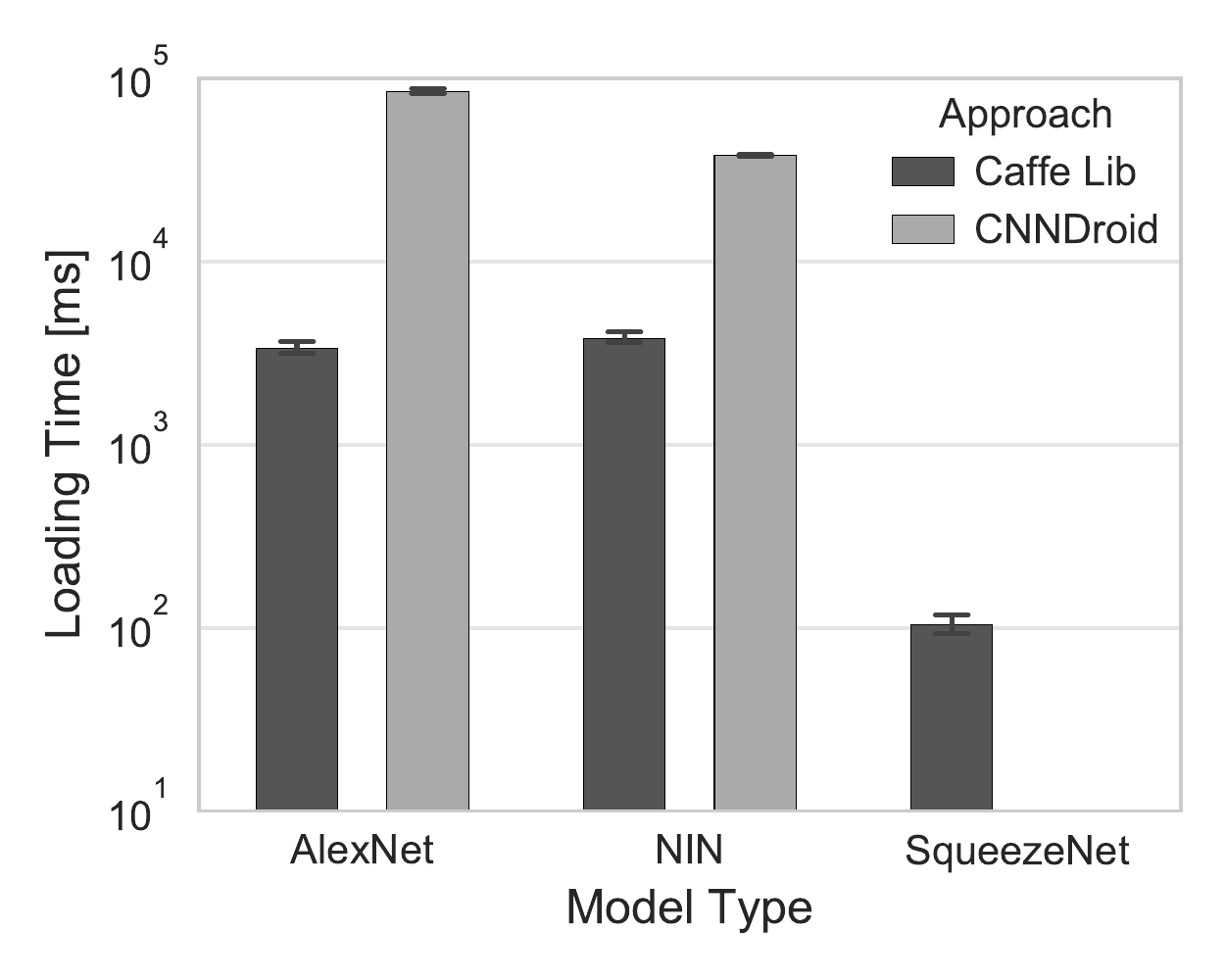}
\caption{Model loading time.}
\label{subfig:model_loading_time}
\end{subfigure}
\hfill
\begin{subfigure}{0.35\textwidth}
\includegraphics[width=\columnwidth]{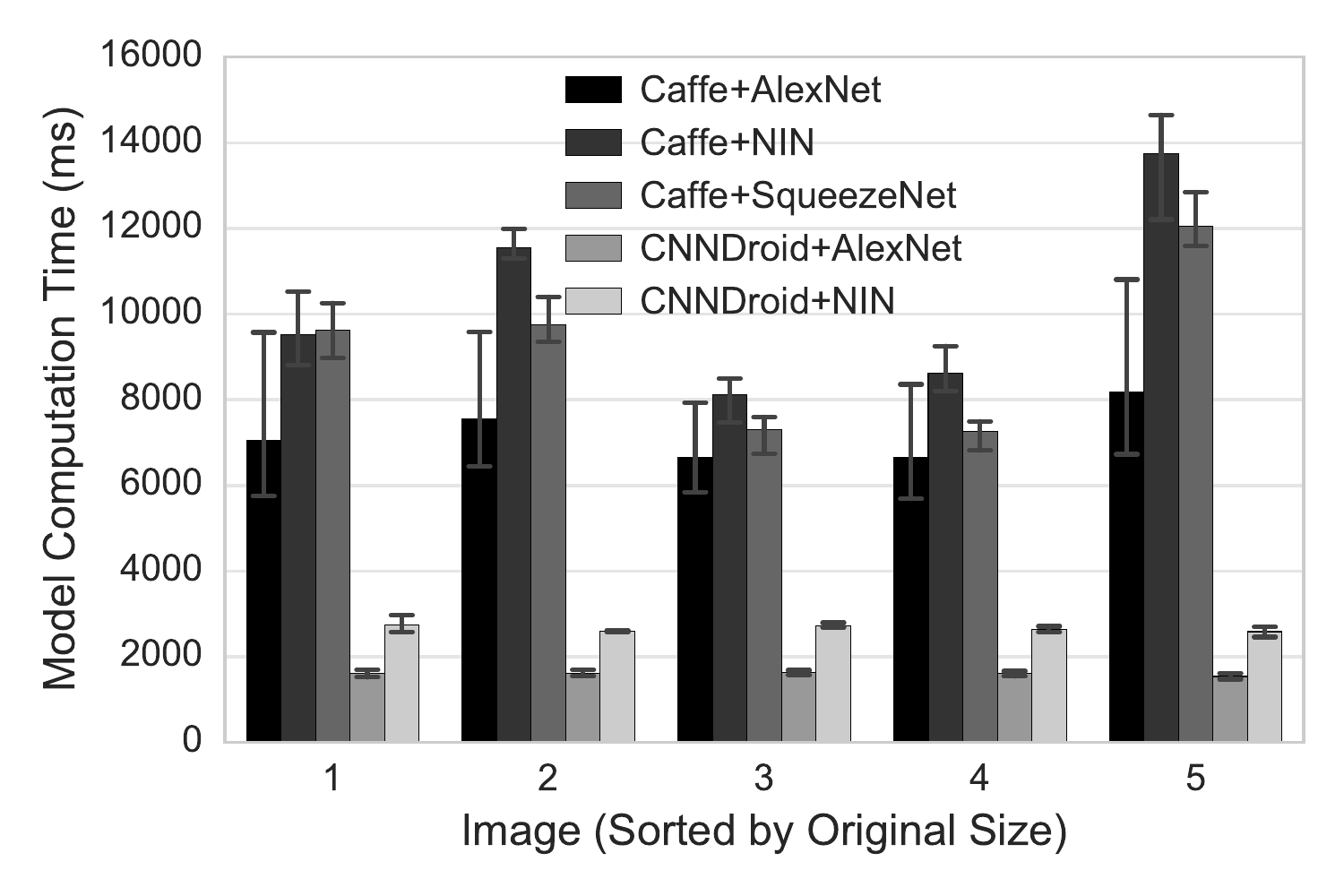}
\caption{Model execution time.}
\label{subfig:model_computation_time}
\end{subfigure}
\hfill
\begin{subfigure}{0.35\textwidth}
\includegraphics[width=\columnwidth]{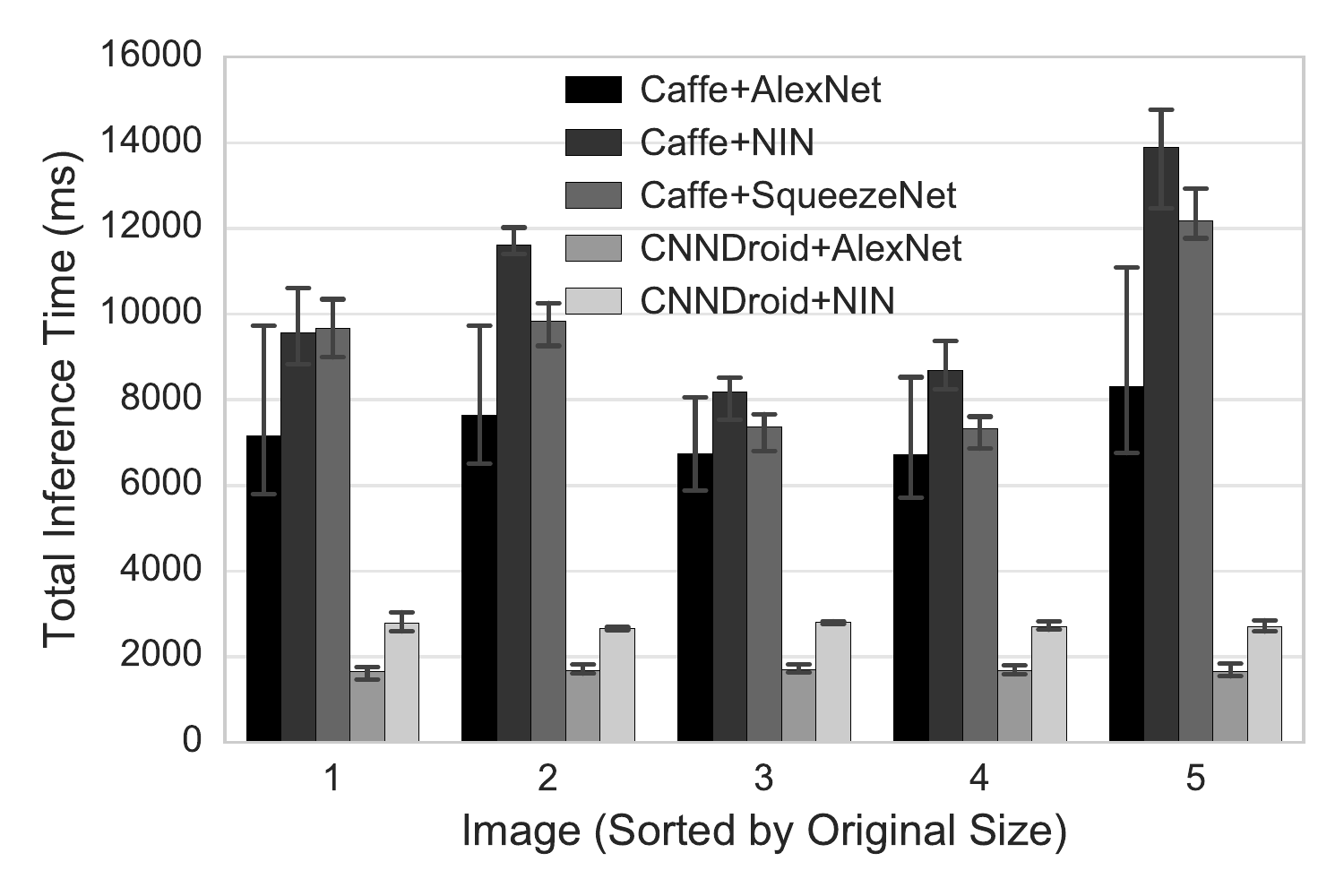}
\caption{End-to-end inference time.}
\label{subfig:model_inference_time}
\end{subfigure}
\caption{\textbf{Caffe on-device inference time.}
We compare two approaches for device-based inference using three CNN models. When using CNNDroid-based approach, trained models need to be converted to supported format.}
\label{fig:dnn_models}
\end{figure*}

\subsection{On-device Inference Performance Analysis}
\label{subsec:ondeviceperf}
We begin by analyzing the performance differences by dissecting the on-device image recognition task with an in-depth study of time breakdown,
and resource utilization.
We focus on understanding the performance of on-device deep inference and the
implications for potential performance improvement. 

\para{Impact of Deep Learning Frameworks.}
The choice of deep learning framework can impact the CNN model design and have a significant impact on the performance,
due to different model sizes and complexities.
We quantify such impacts through both the model loading and inference execution time.
We plot the loading time in Figure~\ref{subfig:model_loading_time}
in \emph{log} scale.
For loading the same model (AlexNet and NIN),
the ported Caffe library takes up to 4.12 seconds,
about 22X faster than using CNNDroid. Furthermore, it only takes an
average of 103.7 ms to load the smallest SqueezeNet model.\footnote{We did not
measure the performance of SqueezeNet using CNNDroid due to the lack of support by CNNDroid.}
This loading happens whenever users first launch the mobile application,
and potentially when a suspended background app is brought back.
Our measurement of CNNDroid's long loading time suggests that users
need to wait for up to 88 seconds to be able to interact with the mobile
app. Although long loading time might be amortized across a number of inference
requests during one user interaction session, it still negatively impacts user
experiences.

Next, we show the time taken to perform inference on the input image using
five different configurations in Figure~\ref{subfig:model_computation_time}.
For each configuration, we measure the computation time taken
for all five images and collect a total of 75 data points.
Each bar represents the average computation time across three versions
of the same image and the standard deviation.
CNNDroid-based AlexNet inference achieves the lowest average of
1541.67 ms, compared to the longest time of 13745.33 ms using
ported Caffe NIN model. Even with the fastest device-based inference,
it still takes three times more than CPU-based cloud inference~\cite{2018ic2e:guo}.
In addition, we plot the end-to-end inference time in Figure~\ref{subfig:model_inference_time}. This total inference time includes the bitmap scaling
time, the GC time, and the model computation time.
CNNDroid-based approach takes an average of
1648.67 ms for performing object recognition on a single image,
about seven times faster than using ported Caffe models.
Based on the response time rules~\cite{ws:nielsenRespTime,ws:mobileappMetrics},
it might lead to poor user experiences when using certain on-device inference. 

\begin{figure}[t]
\centering
\begin{subfigure}[b]{0.24\textwidth}
\includegraphics[width=\columnwidth]{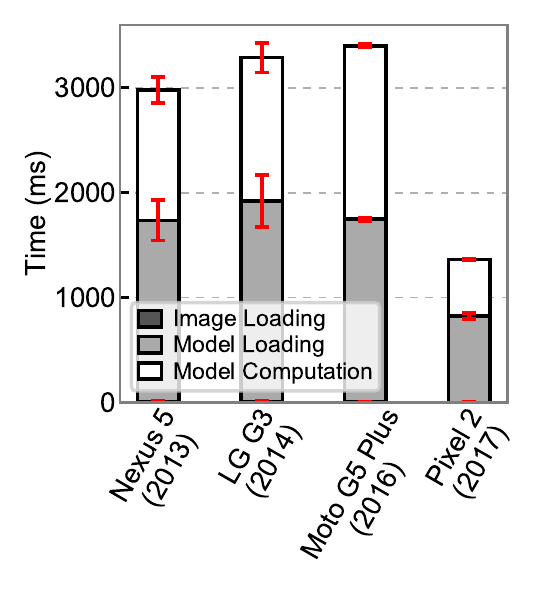}
\caption{\emph{Different mobile capabilities.}}
\label{subfig:device_speeds}
\end{subfigure}
\begin{subfigure}[b]{0.24\textwidth}
\includegraphics[width=\columnwidth]{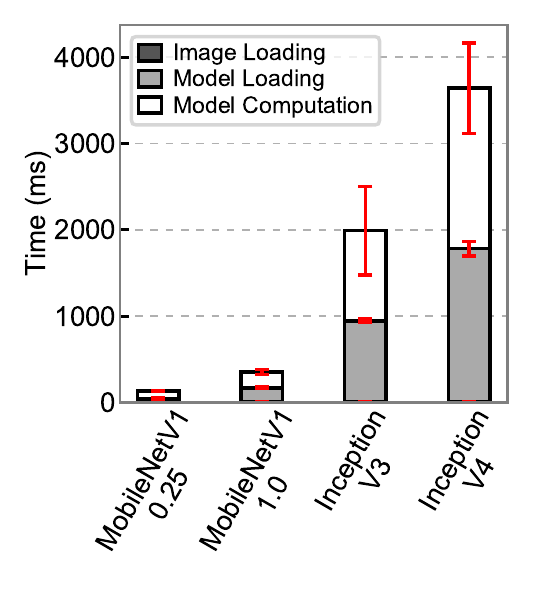}
\caption{\emph{Different CNN models.}}
\label{subfig:model_sizes}
\end{subfigure}
\caption{\textbf{Comparisons of on-device inference using different mobile
devices and CNN models.} The high-end Pixel 2 improves on-device inference
speed by 2.3X compared to other devices. The mobile-specific CNN models,
e.g., \texttt{MobileNet} family, only take an average of 150ms to run and are 27.2x
faster than server-centric models, i.e., \texttt{Inception}. 
Note, the time to load the 330KB image is shown but negligible.}
\label{fig:end_to_end_tensorflow}
\end{figure}

\para{Impact of Heterogenous Mobile Capabilities.} Next, we measure the
end-to-end image classification time for running \texttt{InceptionV3} using
TensorFlow Mobile framework on different mobile devices. In
Figure~\ref{subfig:device_speeds}, we plot the time breakdown of loading the
image and the CNN model, as well as model computation time.  Mobile devices vary
in their ability to load and run the server-centric \texttt{InceptionV3}
model. As we can see, older
or less capable devices (left three bars) take up to 3.5 seconds to return
inference results to mobile users. However, the nearly 2 seconds of model
load time could be amortized across multiple runs or reduced by using smaller
CNN models. Although newer mobile hardware is able to deliver acceptable user performance even for relative complex CNN models, older mobile devices would benefit from having access to simpler CNN models or using cloud-based inferences.

\para{Impact of CNN Models.}
CNN models differ from traditional deep learning models in that they use convolutional layers as their input.
This has two main effects.
First, convolutional layers typically use shared weights for their convolutional layers, decreasing the amount of storage needed for the models themselves.
Second, these convolutional layers increase accuracy for image classification tasks but greatly increase the memory usage of model execution due to the generation of intermediate data.
Therefore we further study the performance differences when running different CNN models on a powerful mobile device.
We choose four popular CNN models and run them using TensorFlow mobile framework on a
Pixel2 mobile phone~\cite{pixel2.tpu}. For each CNN model, we measured the inference time for running \emph{images\_2} test set and calculated the average.  In Figure~\ref{subfig:model_sizes}, we
show that Pixel2 is able to execute both \texttt{MobileNetV1} models in less than 133ms (352ms) on average.
However, Pixel2 takes 48.7X longer to load larger \texttt{Inception}
models and is 5.6X-27.4X slower to perform the model computation using \texttt{Inception}
models. We observed that the model loading time is more consistent compared
to the model computation. This indicates that model computation might be more
subject to the resource interference between foreground and background
applications.

\begin{figure}[t]
\centering
\begin{subfigure}{0.22\textwidth}
\includegraphics[width=\columnwidth]{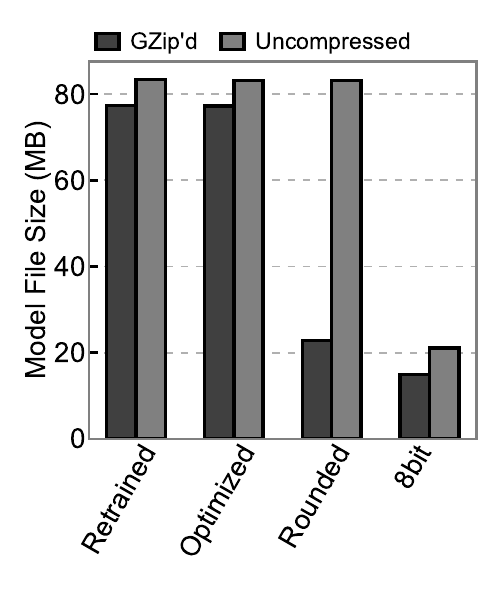}
\caption{\emph{Storage size.}}
\label{subfig:model_sizes}
\end{subfigure}
\centering
\begin{subfigure}{0.26\textwidth}
\includegraphics[width=\columnwidth]{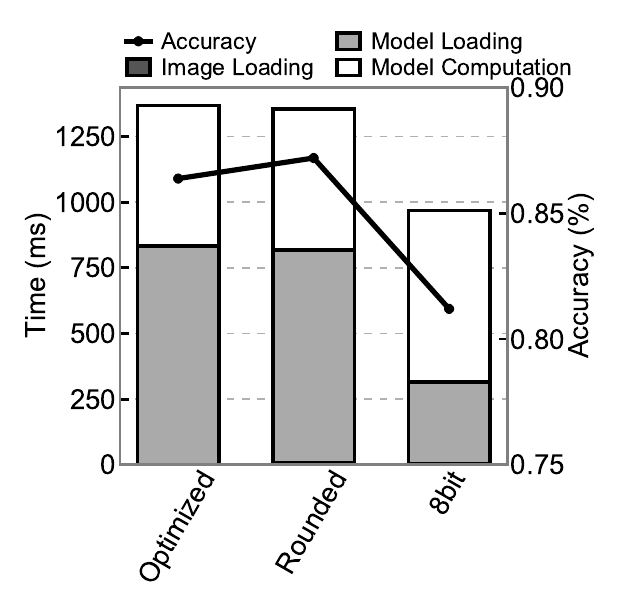}
\caption{\emph{Inference speed and accuracy.}}
\label{subfig:model_speeds}
\end{subfigure}
\caption{\textbf{Comparisons of model compression techniques.}
Model compressions can greatly affect the on-disk storage requirements as well as impacting the accuracy and latency of models.
Overall, the 8-bit quantized model is superior in storage saving and inference speed but experiences 6\% accuracy decrease.
The retrained model is excluded due to unsupported operations on mobile devices.}
\label{fig:compression}
\end{figure}

\para{Impact of CNN Model Compression.} Compactly storing deep learning models is key to our vision of supporting a wide selection of on-device models. However, compression techniques generally trade-off inference accuracy for compression effectiveness. In this section, we first quantify the storage savings of four post-training compression techniques and then compare each technique's impact on inference performance.\par

Figure~\ref{subfig:model_sizes} compares the uncompressed and compressed model sizes. For each model, we plot the baseline uncompressed size (left bar) and the gzip version (right bar). As we can see, the 8-bit quantized model leads to the most storage saving of 75\% regardless of gzip compression. In addition, the unquantized models (retrained and optimized) see only about 7\% savings while the rounding quantized model sees a 72.6\% storage reduction after being gzipped. Our observations suggest that both quantization and gzip compression can lead to significant storage savings, especially when combined.

Next, we compare the inference speed and accuracy of each model.  Figure~\ref{subfig:model_speeds} shows the time taken by each model and its accuracy.  It is important to note that majority of the inference time is in loading the model into memory and thus could be amortized over sequential mobile inferences. However, for one-off mobile inferences, the model loading time dominates the end-to-end inference time.
The small 8-bit quantized model provides the fastest end-to-end response time with the lowest model loading time, but a slightly increased inference time.
In our results, we do see a small accuracy increase for the rounding quantized model. But such observations are not common and would be detected through metadata tracking. In sum, model compression techniques have different impacts on model storage, inference speed and accuracy. We could leverage these observations to carefully select techniques that provide different tradeoffs.

\begin{table}[]
\centering
\caption{\textbf{Summary of GC activities when using CNNDroid-based inference.}
ART uses the default CMS GC, and the GC time takes up to 9.89\%
during model loading, and up to 25\% during user interactions. The average GC pause time can be
up to 39.23 ms.}
\label{tbl:gc_impact}
\resizebox{\columnwidth}{!}{
\begin{tabular}{r|c|rrrr}
\hline
\multicolumn{1}{c|}{\begin{tabular}[c]{@{}c@{}}CNNDroid \\ On-device Inference\end{tabular}} & \cellcolor[HTML]{EFEFEF}\textbf{Phase}                                        & \multicolumn{1}{c}{\cellcolor[HTML]{EFEFEF}\textbf{Duration{[}ms{]}}} & \multicolumn{1}{c}{\cellcolor[HTML]{EFEFEF}\textbf{\begin{tabular}[c]{@{}c@{}}Num. \\ of GC\end{tabular}}} & \multicolumn{1}{c}{\cellcolor[HTML]{EFEFEF}\textbf{\begin{tabular}[c]{@{}c@{}}GC \\ Time {[}ms{]}\end{tabular}}} & \multicolumn{1}{c}{\cellcolor[HTML]{EFEFEF}\textbf{\begin{tabular}[c]{@{}c@{}}GC \\ Pause {[}ms{]}\end{tabular}}} \\ \hline
\cellcolor[HTML]{EFEFEF}AlexNet                                            &                                                                               & 84537.33                                                              & 4.33                                                                                                       & 513.55                                                                                                           & 10.42                                                                                                             \\
\cellcolor[HTML]{EFEFEF}NIN                                                & \multirow{-2}{*}{\begin{tabular}[c]{@{}c@{}}Load \\ Model\end{tabular}}       & 37975                                                                 & 16.67                                                                                                      & 3757.30                                                                                                          & 175.76                                                                                                            \\ \hline
\cellcolor[HTML]{EFEFEF}AlexNet                                            &                                                                               & 11800                                                                 & 4                                                                                                          & 536.55                                                                                                           & 4.60                                                                                                              \\
\cellcolor[HTML]{EFEFEF}NIN                                                & \multirow{-2}{*}{\begin{tabular}[c]{@{}c@{}}User \\ Interaction\end{tabular}} & 17166.67                                                              & 7                                                                                                          & 4307.18                                                                                                          & 274.66                                                                                                            \\ \hline
\end{tabular}}
\end{table}

\begin{figure}[t]
\centering
\includegraphics[width=.4\textwidth]{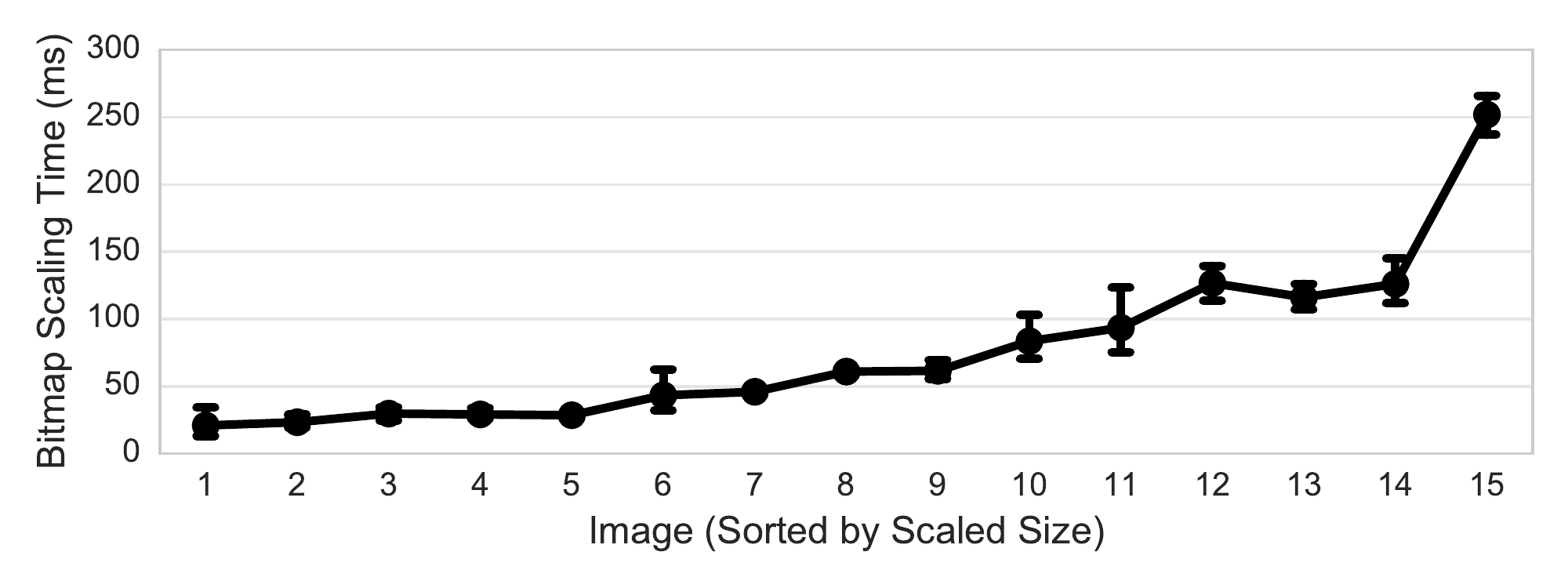}
\caption{
\textbf{Bitmap downscaling time.}
The time taken to downscale the image in the mobile device grow with the image size. Specially, larger images also experiences proportionally longer scaling time because of the limited memory resources assigned to the mobile application.}
\label{fig:scaling_time}
\end{figure}

\para{Impact of Limited Mobile Memory.}
During loading CNNDroid-based models, we observe much more frequent,
and long lasting garbage collecting activities performed by Android Runtime in our mobile device.
When running our app using CNNDroid library, we have to
request for a large heap of 512 MB memory.\footnote{Running
the app with the default 192 MB memory will lead to \texttt{OutOfMemoryError}.}
Even with a large heap, the memory pressure of creating new
objects has lead to a total of 8.33 (and 23.67) GC invocations when using
CNNDroid-based AlexNet (and NIN) model, as shown in Table~\ref{tbl:gc_impact}.
Our evaluation suggests that by allocating more memory to deep learning powered mobile apps,
or running such apps in more powerful mobile devices can mitigate the impact of garbage collection.

\para{Impact of Image Size.}
Because the CNN models in this test only require images of dimension 224 by 224 pixels to perform inference tasks,
we can scale input image to the required dimension before sending.
Figure~\ref{fig:scaling_time} shows the time taken to scale images
with different sizes. Each data point represents the average scaling time
across five different runs.  The time taken to resize image grows as its size increases.
It is only beneficial to downscale an image of size $x_1$ to $x_2$ if
$T_{d} (x_1, x_2) + T_{n} (x_2) \leq T_{n}(x_1)$,
where $T_{d}(x, y)$ represents the time to downscale an image from size $x$ to $y$
and $T_{n}(x)$ denotes the time to upload an image of size $x$ to a cloud server.
For example, based on our measurement, it takes an average of 36.83 ms to upload an image of 172 KB to our cloud server.
Also, from Figure~\ref{fig:scaling_time}, we know that it takes up to 38 ms to resize an image less than 226 KB. By combining these two observations,
it is easy to conclude that directly uploading image one to five is more time efficient. We can expect to make informed decisions about whether
resizing an image of size $x$ before uploading is beneficial or not given enough time measurements of resizing and uploading steps. Our analysis shows that on-device inference's performance bottlenecks mainly exhibit in loading model and computing probability steps.

\begin{figure}[t]
\centering
\begin{subfigure}{0.24\textwidth}
\includegraphics[width=\columnwidth]{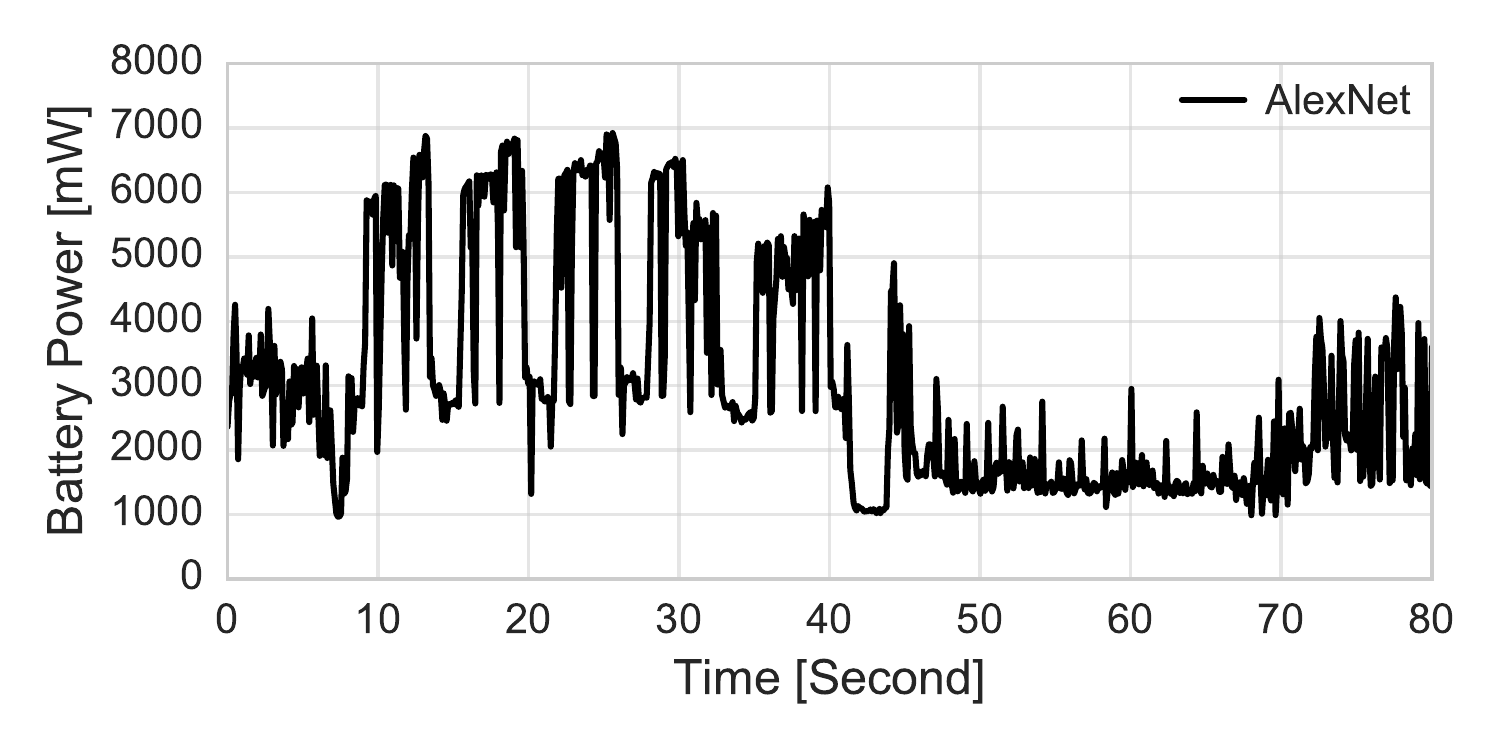}
\caption{\textnormal{Caffe-based energy.}}
\label{subfig:alexnet_battery}
\end{subfigure}
\begin{subfigure}{0.24\textwidth}
\includegraphics[width=\columnwidth]{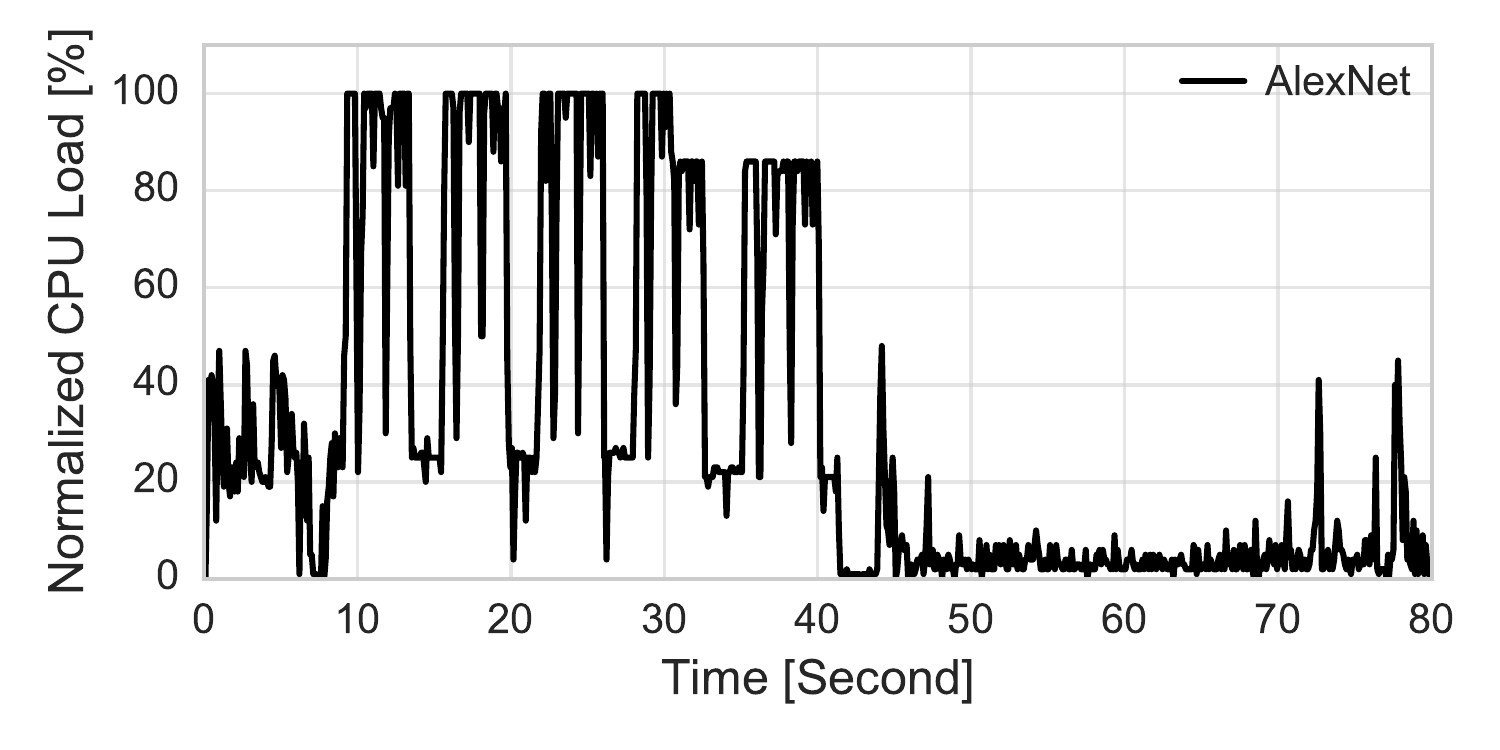}
\caption{\textnormal{Caffe-based CPU.}}
\label{subfig:alexnet_cpu_util}
\end{subfigure}
\begin{subfigure}{0.24\textwidth}
\includegraphics[width=\columnwidth]{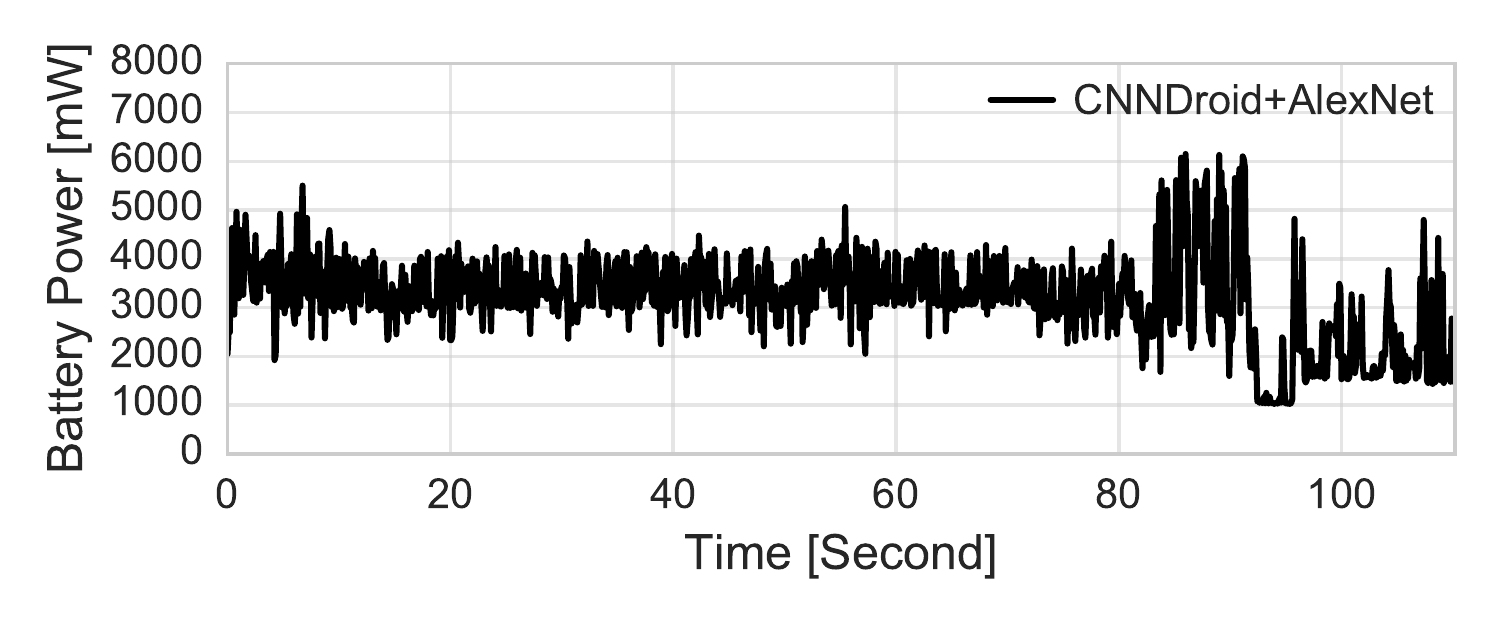}
\caption{\textnormal{CNNDroid-based energy.}}
\label{subfig:cnndroid_alexnet_energy}
\end{subfigure}
\begin{subfigure}{0.24\textwidth}
\includegraphics[width=\columnwidth]{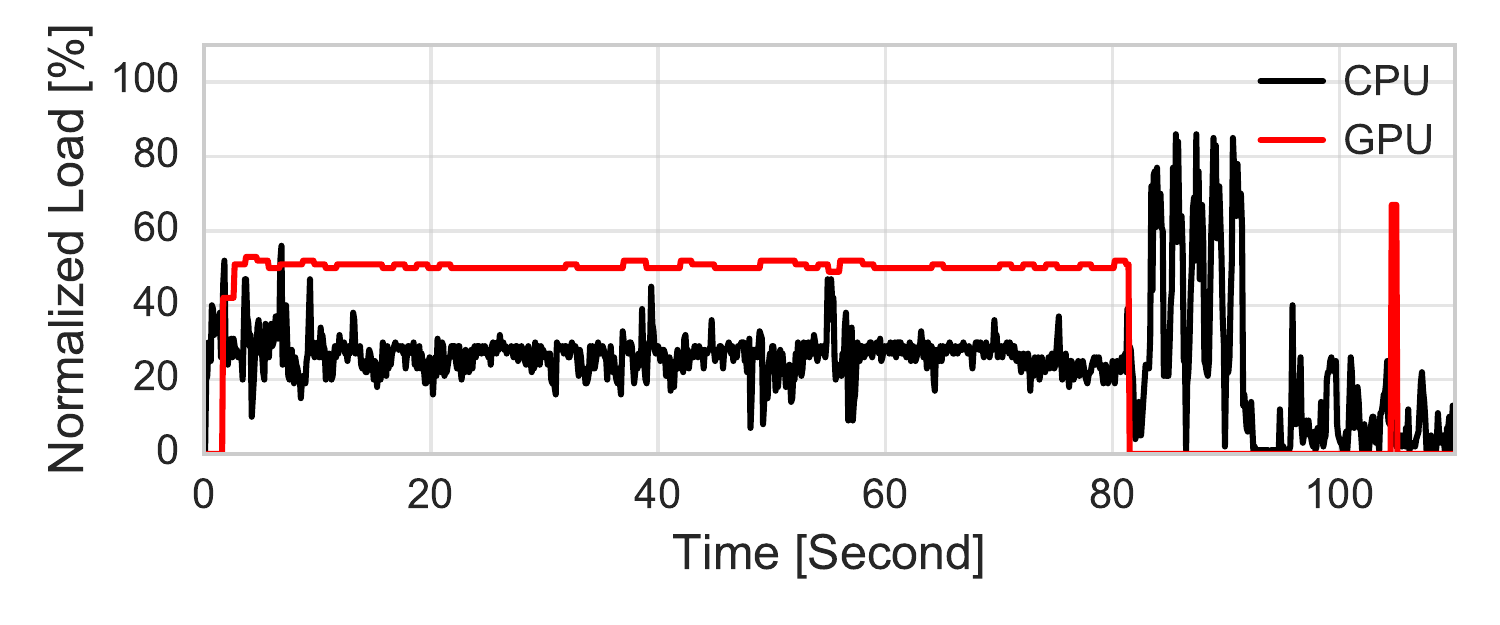}
\caption{\textnormal{CNNDroid-based resource.}}
\label{subfig:cnndroid_alexnet_cpu_gpu}
\end{subfigure}
\caption{
Energy consumption and resource utilization of on-device object recognition task.}
\label{fig:trepn_measurement_caffe}
\end{figure}

\subsection{On-device Inference Resource and Energy Analysis}

In Figure~\ref{fig:trepn_measurement_caffe}, we analyze both the energy consumption and resource utilization when running our app in different configurations . We compare the time-series plots of running AlexNet model using Caffe Android library and CNNDroid framework.
The plots correspond to experiment runs that perform inference tasks on image set one.

For Caffe Android library based approach, we observe an initial energy consumption (and CPU utilization) that increases corresponding to loading AlexNet
CNN model into the memory, a continuation of energy spike during
model computation, and the last phase that corresponds to displaying
images and the most probable label texts, in Figure~\ref{subfig:alexnet_battery} and Figure~\ref{subfig:alexnet_cpu_util}.
The other two CNN models, NIN and SqueezeNet, exhibit very similar usage pattern\footnote{Interested readers can refer to our prior work~\cite{2018ic2e:guo} for additional results of Caffe-based NIN and SqueezeNet models, and CNNDroid-based NIN.}. Specifically, in the case of NIN,
the initial model loading causes the energy consumption to increase from
baseline 1081.24 mW to up to 5000 mW;
when performing the model computation, both the energy consumption
and CPU utilization spikes to more than 7000 mW and 66.2\%.
Note in the case of SqueezeNet, we only observe a very small window of both energy and CPU spikes at the very beginning of measurement. This is because
SqueezeNet can be loaded in 109 ms, compared to more than 3000 ms to load either AlexNet or NIN.

In contrast, we observe two key usage differences in CNNDroid approach, as shown in Figure~\ref{subfig:cnndroid_alexnet_energy} and Figure~\ref{subfig:cnndroid_alexnet_cpu_gpu}. First, CNNDroid-based AlexNet exhibits a longer period of more stable and lower energy consumption
compared to its counterpart in Caffe-based approach.
This is mainly because CNNDroid explicitly expresses some of the data-parallel workload
using \texttt{RenderScript} and is able to offload these workload to more energy-efficient mobile GPU~\cite{ws:gpuEnergy} (indicated by the high GPU utilization during model loading).
Second, the total model computation time is significantly shortened from 40 seconds to around five seconds. In all, by shifting some of computation tasks during
model loading, CNNDroid-based approach successfully reduces the user perceived response time. However, the CNNDroid approach consumes 85.2 mWh energy, over 42\% more than Caffe-based approach. Note 91\% of CNNDroid energy is consumed during model loading phase, and therefore can be amortized by performing inference tasks in batch. In other words, the CNNDroid-based approach is more energy-efficient in performing inference tasks compared to the Caffe-based approach when CNN models are preloaded into the mobile memory.

\section{Understanding Cloud-based Inference}
\label{sec:measure_cloud}

Next, we measured the cloud-based inference performance under different key factors such as cloud server capacity, CNN models, and mobile network. We then analyze the
resource and energy implications of running cloud-based inference. We used four types of cloud-based servers
(summarized in Table~\ref{tab:cloud_servers}) with different capacities. These cloud servers range from burstable servers to GPU-accelerated servers.We run TensorFlow framework behind a flask-based web server and executed inference requests using different images and CNN combinations.

\begin{table}[t]
\centering
\resizebox{0.48\textwidth}{!}{%
\begin{tabular}{@{}r|rrrrr@{}}
\toprule
\textbf{\begin{tabular}[c]{@{}r@{}}cloud \\ server\end{tabular}} & \textbf{vCPU} & \textbf{GPU} & \textbf{\begin{tabular}[c]{@{}r@{}}RAM\\ (GB)\end{tabular}} & \textbf{Storage} & \textbf{Network (Gbps)} \\ \midrule
\textbf{t2.medium} & 2 & N/A & 4 & N/A & N/A \\
\textbf{c5.large} & 2 & N/A & 4 & EBS only & Up to 10 \\
\textbf{p2.xlarge} & 4 & 1 & 61 & N/A & High \\
\textbf{g2.2xlarge} & \begin{tabular}[c]{@{}r@{}}8 Intel\\ Xeon E5-2670\end{tabular} & \begin{tabular}[c]{@{}r@{}}1 Nvidia \\ GRID K520\end{tabular} & 15 & \begin{tabular}[c]{@{}r@{}}60GB \\ instance store\end{tabular} & N/A \\ \bottomrule
\end{tabular}%
}
\caption{\textbf{Cloud servers used in cloud-based measurement.} }
\label{tab:cloud_servers}
\end{table}

\begin{table}[]
\centering
\resizebox{0.45\textwidth}{!}{%
\begin{tabular}{r|rrrr}
\hline
& \multicolumn{4}{r}{\textbf{image recognition time breakdown (ms)}} \\ \cline{2-5}
\multirow{-2}{*}{\textbf{inference mode}} & \begin{tabular}[c]{@{}r@{}}model \\ loading\end{tabular} & \begin{tabular}[c]{@{}r@{}}image \\ resizing\end{tabular} & \begin{tabular}[c]{@{}r@{}}image \\ uploading\end{tabular} & \begin{tabular}[c]{@{}r@{}}probability\\ computing\end{tabular} \\ \hline
g2.2xlarge+CPU & N/A & 76.2 & 36.8 & 238.6 \\
g2.2xlarge+GPU & N/A & 76.2 & 36.8 & 18.6 \\
{\color[HTML]{9B9B9B} Nexus5+Caffe} & {\color[HTML]{9B9B9B} 2422.1} & {\color[HTML]{9B9B9B} 80.0} & {\color[HTML]{9B9B9B} N/A} & {\color[HTML]{9B9B9B} 8910.6} \\
{\color[HTML]{9B9B9B} Nexus5+CNNDroid} & {\color[HTML]{9B9B9B} 61256.2} & {\color[HTML]{9B9B9B} 70.4} & {\color[HTML]{9B9B9B} N/A} & {\color[HTML]{9B9B9B} 2131.7} \\ \hline
\end{tabular}%
}
\caption{\textbf{Impact of hardware acceleration.} We measured the image recognition time with and without using GPU. We also included the on-device performance from Figure~\ref{fig:dnn_models} as baselines.}
\label{tab:hardware_acceleration}
\end{table}

\subsection{Cloud-based Inference Performance Analysis}
In this section, we present the measurement results of image recognition time and mobile resource utilization for both cloud-based and on-device inference.

\begin{figure}[t]
\centering
\includegraphics[width=0.4\textwidth]{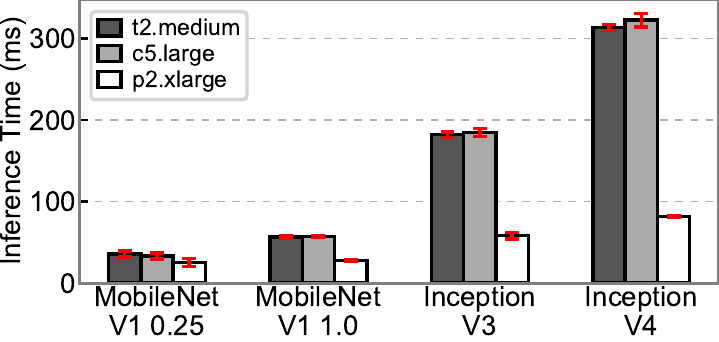}
\caption{\textbf{Comparison of inference latency of cloud-based inference
with four CNN models.} The \texttt{InceptionV4} that runs on the \texttt{p2.xlarge} GPU-accelerated server is over 2.5X faster
than the \texttt{MobileNet 0.25} on the MotoX.
This highlights the advantage of cloud-based inference in providing high-accuracy and low-latency results and demonstrates the potential of cloud-based inference for enabling inference latency and accuracy trade-offs.}
\label{fig:incloud_model_runtime}
\end{figure}

\para{Impact of hardware acceleration.}
Table~\ref{tab:hardware_acceleration} summarizes the average end-to-end performance and resource consumption of executing object recognition using both cloud-based and on-device inference modes.
We use CPU-only Caffe framework and GPU accelerated Caffe framework for toggling the CPU and GPU mode in our \texttt{g2.2xlarge} server. 
For each inference mode, we repeat the recognition tasks using all fifteen images and three CNN models. We measure the time to execute each step and calculate the average.
Similar to on-device inference the task of image recognition is further broken down into four steps: loading CNN models into memory, downscaling image input to desired dimension, uploading input data to the cloud server, and executing the inference.
In contrast to on-device deep inference, in the cloud-based scenario the time to load models is negligible because models already reside inside the memory and can be used to execute the inference task immediately.
Similarly, on-device mode does not incur any time for uploading image bitmaps.

Recall, the inference time is the sum of rescaling, uploading bitmap and inference execution over one bitmap, and the recognition time is the sum of amortized model loading time over a batch of images and inference time.
The average cloud-based inference time is 351.59 ms/131.59 ms when using CPU-only/GPU of a well-provisioned cloud instance hosted in a nearby data center.
As shown, because inference tasks are typically data-parallel and therefore can be accelerated by up to 10x when using GPU.
However, we should note that such results represent a lower bound performance of real-world setting.
In a real-world deployment scenario, image recognition time can last much longer due to reasons such as overloaded cloud servers and variable mobile network conditions.
The total inference time when running on-device is almost 9 seconds when using Caffe-based model, and 2.2 seconds when using CNNDroid model.

\para{Impact of model startup latency.} Next, we studied the inference execution
time difference of \emph{hot start} versus \emph{cold start} model. Here, hot
start inference time refers to the time taken if the CNN model has been loaded
into the memory, i.e., the model has been used to service inference requests
previously. In the case of cold start, we reloaded the CNN model every time
before executing the inference request and measured the cold start inference
time starting from loading the CNN model until the inference response is
generated.

Table~\ref{eval:model-summary} summarizes the CNN model accuracies and inference execution time averaged across 1000 inference requests. First, we observe a distinct correlation between accuracy and inference time. Additionally, the cold start time is generally much larger than the hot start
time, and increases at a much faster rate. Another important observation is
that cold start time is harder to predict than hot start time. For example, in
the case of DensNet, we observed about 23X increase compared to 12X increase for a similarly sized MobileNet model. The increased time can be attributed to complex CNN structure which requires longer setup time and large CNN model size that leads to longer time to load into GPU memory. Therefore it is critical to keep important and often used CNN models in the memory.

\begin{table}[t]
\centering
\resizebox{0.48\textwidth}{!}{%
\begin{tabular}{r|rr|rr}
\hline
& \multicolumn{2}{r|}{\textbf{accuracy (\%)}} & \multicolumn{2}{r}{\textbf{inference time (ms)}} \\ \cline{2-5}
\multirow{-2}{*}{\textbf{\begin{tabular}[c]{@{}r@{}}CNN \\ model\end{tabular}}} & top1 & top5 & \multicolumn{1}{r|}{hot start} & cold start \\ \hline
SqueezeNet & 49.0 & 72.9 & \multicolumn{1}{r|}{28.61 $\pm$ 1.13} & 173.38 $\pm$ 25.73 \\
MobileNetV1 0.25 & 49.7 & 74.1 & \multicolumn{1}{r|}{25.73 $\pm$ 1.22} & 272.81 $\pm$ 45.00 \\
MobileNetV1 0.5 & 63.2 & 84.9 & \multicolumn{1}{r|}{26.34 $\pm$ 1.19} & 302.77 $\pm$ 45.50 \\
DenseNet & 64.2 & 85.6 & \multicolumn{1}{r|}{49.55 $\pm$ 3.21} & 1149.04 $\pm$ 108.00 \\
MobileNetV1 0.75 & 68.3 & 88.1 & \multicolumn{1}{r|}{28.02 $\pm$ 1.14} & 351.92 $\pm$ 47.38 \\
MobileNetV1 1.0 & 71.8 & 90.6 & \multicolumn{1}{r|}{28.15 $\pm$ 1.22} & 421.23 $\pm$ 47.14 \\
NasNet Mobile & 73.9 & 91.5 & \multicolumn{1}{r|}{55.31 $\pm$ 4.09} & 2817.25 $\pm$ 123.73 \\
InceptionResNetV2 & 77.5 & 94.0 & \multicolumn{1}{r|}{76.30 $\pm$ 5.74} & 2844.29 $\pm$ 106.49 \\
InceptionV3 & 77.9 & 93.8 & \multicolumn{1}{r|}{55.75 $\pm$ 1.20} & 1950.71 $\pm$ 101.21 \\
InceptionV4 & 80.1 & 95.1 & \multicolumn{1}{r|}{82.78 $\pm$ 0.89} & 3162.24 $\pm$ 133.99 \\
NasNet Large & 82.6 & 96.1 & \multicolumn{1}{r|}{112.61 $\pm$ 6.09} &
7054.52 $\pm$ 238.36 \\
\hline
\end{tabular}%
}
\caption{\textbf{Summaries of CNN model statistics through empirical
measurement.} We measured the average inference time (with standard deviation)
with/out startup latency for each model running on an EC2
\texttt{p2.xlarge} GPU server. }
\label{eval:model-summary}
\end{table}

\para{Impact of CNN models.} We next study how different CNN models and cloud
servers lead to different inference execution time. We evaluated four different types of CNN models with reasonable inference accuracy (more detailed in Table~\ref{eval:model-summary}).
For each CNN model, we used an image of 110KB and repeated the measurement for 1000 times.
We measured the inference execution time, which is defined as the time to generate output labels based on input images.
In Figure~\ref{fig:incloud_model_runtime}, we plotted the average inference execution time for running these four CNN models in three cloud servers with different capacities.
We make a few observations.
First, regardless of the cloud servers in use---whether the cloud server is a
burstable \texttt{t2.medium} that might be subject to performance
fluctuations~\cite{Wang:2017:UBI:3107080.3084448} or a powerful GPU server \texttt{p2.xlarge}, the
simplest of the four CNN models \texttt{MobileNetV1 0.25} takes less than 50 milliseconds
with up to 30\% difference between \texttt{p2.xlarge} and \texttt{t2.medium}.
Second, as the CNN models become more computation-intensive, both CPU servers start to take more than 313.7 milliseconds to finish in the case of \texttt{InceptionV4}.
Meanwhile, the GPU server \texttt{p2.xlarge} was able to deliver low execution time in less than 82 milliseconds, which is still over 2.5X faster than running the simplest CNN model \texttt{MobileNetV1 0.25} on the MotoX mobile phone.
Our measurements suggest that CNN models with different computation complexities, in addition to cloud servers, take different amount of execution time. These results also demonstrate the potential of using cloud-based inference for making inference latency and accuracy trade-offs dynamically.

\begin{figure}[t]
\centering
\begin{subfigure}{0.24\textwidth}
\includegraphics[width=\textwidth]{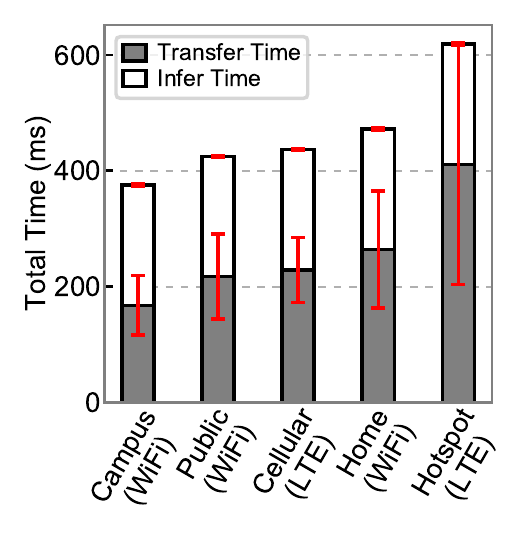}
\caption{Cloud-based inference response time.}
\label{edge_end_to_end}
\end{subfigure}
\begin{subfigure}{0.24\textwidth}
\includegraphics[width=\textwidth]{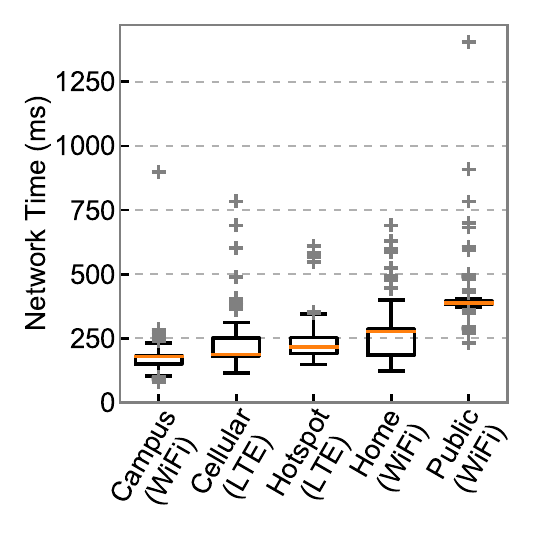}
\caption{Mobile network latency.}
\label{edge_mobile_network}
\end{subfigure}
\caption{\textbf{Comparisons of cloud-based inference under different mobile network conditions.}
Under poor network connectivity, such as when using cellular hotspot, the transfer time
almost doubled when compared to university WiFi. In addition, we demonstrate that network conditions play a critical role in cloud-based inference.
}
\label{fig:edge_inference_perf}
\end{figure}

\para{Impact of mobile network conditions.} We evaluate the end-to-end classification time when using models hosted in edge servers with the optimized model pre-loaded. Figure~\ref{fig:edge_inference_perf} shows the average classification time using an edge server in a variety of network conditions. The majority of the classification time is network transfer time with up to 66.7\% in the cellular hotspot case. Overall, the edge-server based classification ranges from 375ms to 600ms in a well-provisioned \texttt{t2.medium} cloud server. When running the same classification task the Pixel 2 takes 536ms with a preloaded model and is on par with edge based inference. Mobile-based inference only delivers acceptable performance for newer and high-end mobile devices while edge-based inference is a viable option even under poor network condition. \sysname can leverage this observation to dynamically select inference locations.

\begin{table}[]
\centering
\resizebox{0.42\textwidth}{!}{%
\begin{tabular}{r|rrrr}
\hline
& \multicolumn{4}{r}{\textbf{mobile resource consumption}} \\ \cline{2-5}
\multirow{-2}{*}{\textbf{inference mode}} & \begin{tabular}[c]{@{}r@{}}CPU\\ (\%)\end{tabular} & \begin{tabular}[c]{@{}r@{}}GPU\\ (\%)\end{tabular} & \begin{tabular}[c]{@{}r@{}}Mem\\ (GB)\end{tabular} & \begin{tabular}[c]{@{}r@{}}battery\\ (mW)\end{tabular} \\ \hline
g2.2xlarge+CPU & 6.2 & 0.9 & 1.28 & 1561.6 \\
g2.2xlarge+GPU & 6.4 & 0.4 & 1.31 & 1560.2 \\
{\color[HTML]{9B9B9B} Nexus5+Caffe} & {\color[HTML]{9B9B9B} 35.0} & {\color[HTML]{9B9B9B} 80.0} & {\color[HTML]{9B9B9B} 1.64} & {\color[HTML]{9B9B9B} 3249.0} \\
{\color[HTML]{9B9B9B} Nexus5+CNNDroid} & {\color[HTML]{9B9B9B} 22.2} & {\color[HTML]{9B9B9B} 70.4} & {\color[HTML]{9B9B9B} 1.75} & {\color[HTML]{9B9B9B} 2962.6} \\ \hline
\end{tabular}%
}
\caption{\textbf{Mobile resource consumption.} We measured the mobile resource consumption when using cloud-based inference. We also included the on-device performance from Figure~\ref{fig:dnn_models} as baselines.}
\label{tab:mobile_resource_consumption}
\end{table}

\subsection{Cloud-based Inference Resource and Energy Analysis}
Table~\ref{tab:mobile_resource_consumption} shows the resource consumption of mobile device when running the object recognition mobile application.
As a baseline, we measure the performance when the device is idle and the screen is turned on. The CPU utilization and power consumption is 3.17\% and 1081.24 mW respectively. Cloud-based mode consumes roughly the same amount of CPU and 
44.4\% more power consumption comparing to the baseline. However, on-device mode 
not only incurs significantly higher CPU utilization (and in the case of CNNDroid, GPU utilization as well), but also require two times more power consumption when compared to the baseline.
In all, we can calculate the energy consumption of different inference modes by multiplying the average inference (recognition) time by the average power consumption. Cloud-based inference requires as low as  
0.057 mWh energy when using faster GPU computation, and on-device based inference consumes up to 
8.11 mWh.

In sum, cloud-based inference exhibits substantial benefits in terms of inference response time and mobile energy savings over on-device inference, in this case by two orders of magnitude. This is due to more powerful processing power, shorter durations of inference and efficient use of network interfaces.

\section{Managing and Selecting CNN models}
\label{sec:design}

In this section, we describe our \emph{multi-models} approach, \modiselect, to mitigating the impact of the mobile network variations on cloud-based inference performance. \modiselect manages a set of CNN models that exhibit
different execution time and accuracy trade-offs, and selects the bests CNN
model for a given mobile inference request.
The key insight is that the variations of transferring the input data for an inference request can be masked with CNN models that take differing amounts of time to execute.
When considering which CNN model to use, we assume that all CNN models are already loaded into the memory and that CNN model performance profiles are measured and managed by individual inference servers.
We further assume that image preprocessing is handled in an intelligent way by the mobile device.
In Table~\ref{tbl:symbols}, we list all the symbols used in this paper.

\begin{table*}[t]
\begin{minipage}[b]{0.64\textwidth}
\centering
\resizebox{\columnwidth}{!}{%
\begin{tabular}{l|l}
\textbf{Symbol}          & \textbf{Meaning}                                            \\ \hline
\rowcolor{Gray}
$T_{sla}$, $T_{start}$, $T_{budget}$      & Response time SLA, start time, and
remaining name of a mobile inference request.                               \\
\rowcolor{Gray}
$T_{threshold}$ & Confidence threshold of inference performance.     \\
\rowcolor{Gray}
$T_{input}$, $T_{output}$, $T_{nw}$     & Nework time to send inference request and
response, and both. \\
\rowcolor{Gray}
$T_{D}$         & Expected on-device inference time.                 \\
\rowcolor{Gray}
$T_{U}$, $T_{L}$         & The soft and hard time budgets.\\
\rowcolor{Gray}
$T_{R}$, $T_{E}$         & The time budget and exploration ranges.   \\
$K$             & The total number of models.                        \\
$M_E$           & The exploration set of models.                     \\
$\mu(m)$, $\sigma(m)$  & Average and standard deviation of historical inference time of model $m$.               \\
$\bm{Pr}(m)$    & Probability of model $m$ for performing inference. \\
$\bm{A}(m)$, $\bm{T}(m)$, $\bm{U}(m)$      & Accuracy, inference time and utility of CNN model $m$.                             \\
\end{tabular}
}
\caption{\textbf{Symbols summary.} Shaded symbols are specific to
inference requests from different mobile devices, while others
are related to CNN models.}
\label{tbl:symbols}
\end{minipage}
\hfill
\begin{minipage}[b]{0.34\textwidth}
\centering
\includegraphics[width=\textwidth]{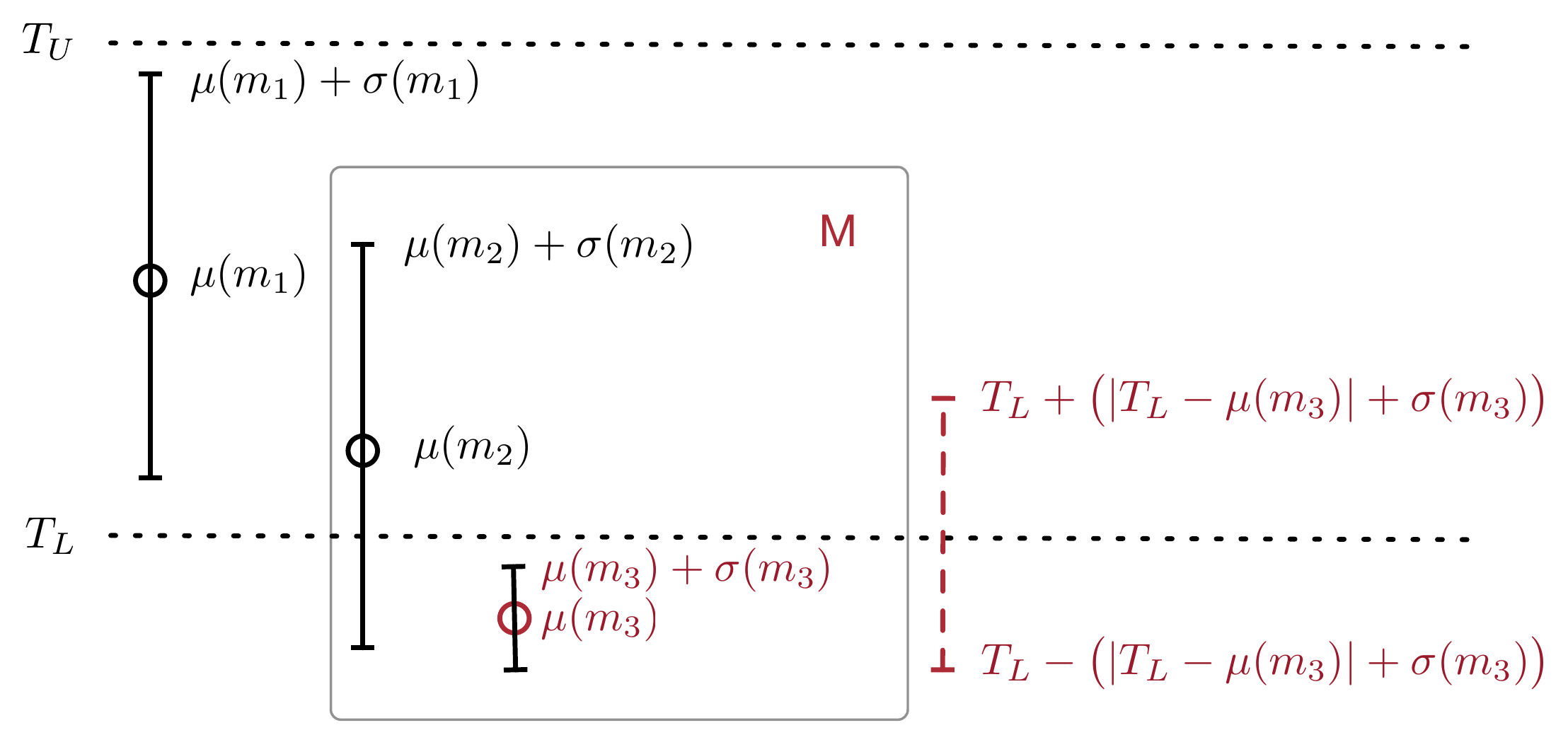}
\captionof{figure}{\textbf{Walkthrough of \modiselect.} Here, \modiselect has access to CNN model performance profiles
and inference accuracy: $\bm{A}(m_3) > \bm{A}(m_1) > \bm{A}(m_2)$.}
\label{fig:design.model.selection}
\end{minipage}
\end{table*}

For a mobile inference request, we assume that developers provide the target
response time that indicates how long the end-to-end inference should take.
This target time can alternatively be expressed as Service Level Agreement between mobile developers and cloud-based inference service providers.
Without loss of generality, we refer to this target response time as $T_{sla}$.

Next, \modiselect estimates the remaining time $T_{budget}$ that an inference
request should finish executing. $T_{budget}$ is calculated by taking the difference between $T_{sla}$ and the network transfer time $T_{nw}$. $T_{nw}$ can be estimated conservatively with $2*T_{input}$ where $T_{input}$ denotes the time taken to send input data from the mobile device to the cloud-based inference server. In our application scenarios, we could expect $T_{input} \geq T_{output}$ given inference requests, e.g., images, are often larger than inference responses, e.g., text labels. To summarize, $T_{budget}$ can be calculated as: $T_{budget}=T_{sla} - 2*T_{input}$.

To take into account of the CNN performance profiles getting outdated---leading to less accurate estimation of execution time, we define a threshold $T_{threshold}$ which indicates how uncertain we are about the model performance profiles. The larger the value of $T_{threshold}$, the more outdated the inference performance profiles. In order to effectively explore all potentially high-accuracy models without violating SLA, we then expand the notation of time budget $T_{budget}$ to a range $T_R = [T_L, T_U]$, where  $T_U=T_{budget}$ and $T_L  = T_U - T_{threshold}$ (see Figure~\ref{fig:design.model.selection}).  Intuitively, $T_U$ represents the maximum amount of time that \modiselect can use for generating an inference response without risking SLA violations. We refer to $T_L$ as the \emph{hard time limit}. On the other hand, $T_U$ is referred to as the \emph{soft time limit} and provides \modiselect the flexibility to explore a subset of high-accuracy models $M_{E}$ that exhibit different execution time $\{\bm{T}(m) | \forall m \in M_{E}\}$.

Currently, $T_{threshold}$ is configured by \sysname's users, e.g., mobile
application developer, as any values in the range of $[0, T_{D}]$, where $T_{D}$
represents the expected on-device inference time. We choose to bound
$T_{threshold}$ this way to restrict the exploration set $M_{E}$ and mitigate  the undesirable behavior of starting on-device inference prematurely when cloud-based inference can finish without violating the SLA. However, \sysname could also dynamically adjust $T_{threshold}$ based on its confidences of model performance profiles and will be explored as part of future work.

\subsection{Opportunistic Model Selection}

Next, we describe in detail how \sysname utilizes both the model performance profiles and the time budget range $T_R$ to first pick a base model, then construct a set of eligible models $M_E$ that is worth exploring, and last probabilistically select the model for executing the inference request.
Our three-staged algorithm is designed to gradually improve our estimation of model performance profiles without incurring additional profiling overhead.
In addition, if we are under time pressure to select models, our algorithm could be stopped any time after the first stage and will still select a quality model for performing the inference.
In Figure~\ref{fig:design.model.selection} we provide an example walkthrough of
\modiselect.

\para{Stage one: greedily picking the baseline model.}
In this stage, \sysname takes all the existing models and selects a base model
$m_j$ as follows.

\begin{alignat}{4}
& \underset{j}{\text{maximize}} & \hspace{5mm}
&  \bm{A} (m_j)  \label{eq:obj}\\
& \text{subject to} && \mu(m_j) + \sigma(m_j) < T_U , \; j = 1\ldots K.
\label{eq:c1} \\
&&& \mu(m_j) - \sigma(m_j) < T_L , \; j = 1 \ldots K.  \label{eq:c2}
\end{alignat}

The high level idea is to select the most accurate model Equation~\eqref{eq:obj}
that are likely to finish execution within specified SLA target
Equation~\eqref{eq:c1} without triggering on-device inference
Equation~\eqref{eq:c2}.
Given that cloud-based inference execution might experience performance
fluctuations that lead to a wider inference execution
distribution~\cite{Sharma:2016:CVM:2988336.2988337,8298516}, we take into
account of the standard deviation of model inference time and only select models
that satisfy both the soft time limit $T_L$ and the hard time limit $T_U$. By
doing so, the selected models are of high accuracy and are very likely to finish
execution within specified SLA.
In the example walkthrough in
Figure~\ref{fig:design.model.selection}, \modiselect will select model $m_3$ as the
base model. In scenarios where there is no base CNN models that satisfy the
constraints, e.g., variability of mobile networks, \modiselect chooses the CNN model with the lowest average inference
time $\mu(m_j)$ in order to provide a best-effort at SLA attainment.

\para{Stage two: optimistically constructing the eligible model set.} To account
for cloud-based inference variations, either due to workload spikes~\cite{Bodik:2010:CMG:1807128.1807166} or
insufficient/outdated CNN model performance profiles, \sysname leverages the
basic idea of exploiting and exploration~\cite{Yogeswaran2012}.
Given the
base CNN model $m^*$, we then explore other potential CNN models that should
also be
inside the exploration set $M_E$. To do so, we leverage $m^*$ performance
profile and construct the exploration range $T_E$.

Specifically, we expand the
hard time limit $T_L$ with $\mu(m^*)$ and $\sigma(m^*)$ to construct the
exploration range $T_E$ with an acceptable distance $(|T_L - \mu(m^*)|) +
\sigma(m^*)$ as below.

\begin{equation}
T_E=
\begin{cases}
[\mu(m') + \sigma(m'),
2T_L - \mu(m') + \sigma(m')], \text{if}\ T_L > \mu(m') \\
[2T_L - \mu(m') + \sigma(m'), \mu(m') + \sigma(m')], \text{otherwise.}
\end{cases}
\notag
\end{equation}

Given this, we
construct the exploration CNN model set $M_E = \{m \mid \mu(m) \in T_E \text{
and } \mu(m) +
\sigma(m) < T_U \}$. As such, all CNN models in $M_E$ satisfy our target
performance while providing \sysname the opportunity to exploit the trade-offs
between inference accuracy and time. In Figure~\ref{fig:design.model.selection},
only model $m_2$ and $m_3$ are marked as the members of $M_E$.

\para{Stage three: opportunistically selecting the CNN model.} \sysname selects
the CNN model $m'$ that balances the risk of SLA violations and the exploration
reward. Concretely, we calculate the utility for each CNN model $\bm{U}(m)$
based on its inference accuracy and the likelihood to violate response time SLA.

\begin{equation}
\bm{U}(m) = \bm{A}(m) \frac{T_U - \big(\mu(m) + \sigma(m)\big)}{| T_L - \mu(m) |}\label{eq:utility}
\end{equation}

\modiselect than calculates the selection probability $\bm{Pr}(m) =
\frac{1}{\sum\limits_{n \in \bm{M_E}} \bm{U}(n)} \bm{U}(m)$ and picks the $m'$
accordingly. This helps avoids choosing CNN models with lower inference accuracy, wider inference time
distribution, and outdated performance profile.

\subsection{Experimental Evaluation of \modiselect}
\label{sec:eval}

We quantify the effectiveness of \modiselect, in dynamically selecting the most
appropriate CNN model to avoid missing target response time while achieving good
inference accuracy. We use a mix of experiments and simulations with real-world
CNN models.
We run our experiments using an Virginia-based Amazon EC2 \texttt{p2.xlarge} GPU
server that manages two retrained CNN models, \texttt{MobileNetV1 0.25} and
\texttt{InceptionV3}. After warming up the inference server, we have our
Massachusetts-based image recognition Android application on MotoX (late 2017)
that send inference requests of preprocessed images (average 330KB) over campus
WiFi. For each SLA target, our mobile application sends 1000 inference requests
and measure both the inference accuracy and end-to-end inference time.
For our simulations, we leverage a number of CNN models,
summarized in Table~\ref{eval:model-summary}\cite{tensorflow_models,
DBLP:journals/corr/HowardZCKWWAA17,DBLP:journals/corr/SzegedyVISW15,
NIPS2012_4824,DBLP:journals/corr/IandolaMAHDK16,DBLP:journals/corr/SimonyanZ14a,resnet}, that expose different accuracy and
inference time trade-offs. Our simulations are seeded with empirical
measurements of CNN model execution time and mobile network conditions.
For each simulation, we generate 10,000 inference requests with a predefined SLA target
and record the model selected by
\modiselect (and baseline algorithms) and relevant performance metrics. We
repeat each simulation for different SLA target and network profiles
combination.

\begin{figure}[t]
\centering
\includegraphics[width=0.45\textwidth]{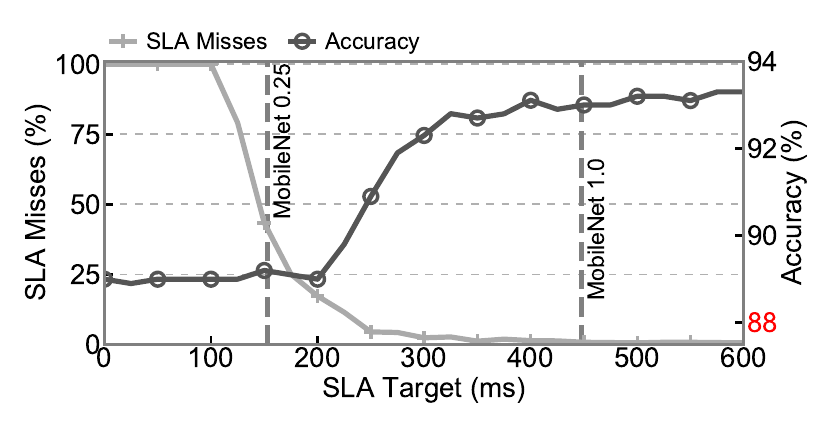}
\caption{\textbf{End-to-end performance of \modiselect.} \modiselect automatically
transitions between two CNN models as the target SLA increases. This increased budget allows \modiselect to use the higher latency but more CNN model.}
\label{eval:subfig:prototype:trade-off}
\end{figure}

\begin{figure}[t]
\centering
\begin{subfigure}{0.44\textwidth}
\centering
\includegraphics[width=\columnwidth]{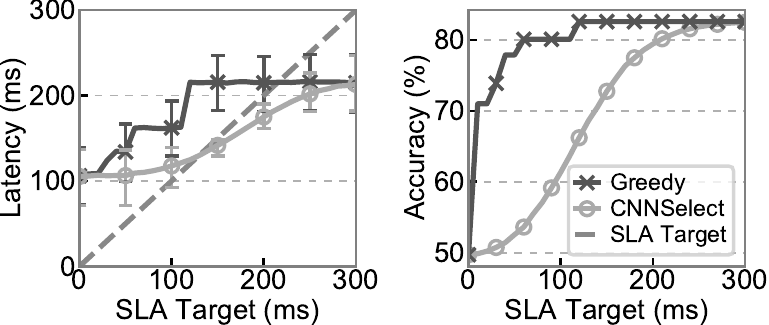}
\caption{Comparison of average end-to-end latency (error bars show 25th and 75th percentiles) and accuracy. }
\label{eval:subfig:two-models:compare}
\end{subfigure}
\begin{subfigure}{0.45\textwidth}
\centering
\includegraphics[width=\columnwidth]{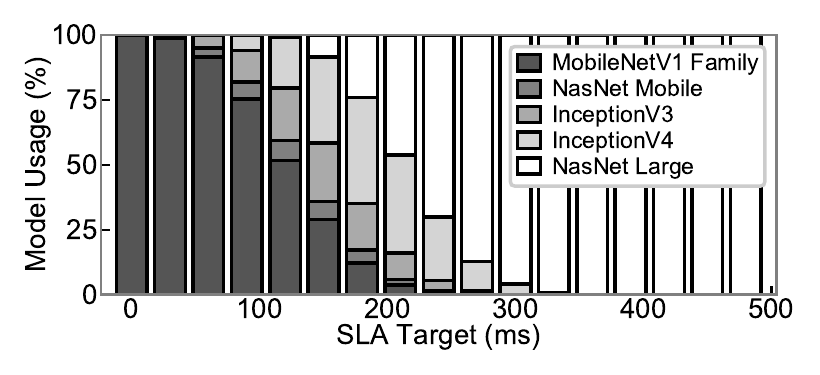}
\caption{\emph{CNN models selected by \modiselect.}  As the SLA increases the time budget for inference increases and \modiselect chooses more accurate models. }
\label{eval:subfig:two-models:model-usage}
\end{subfigure}
\caption{\textbf{Comparison of \modiselect to the \emph{greedy}
algorithm}.
\modiselect tracks the time left to the SLA target allowing it to meet the SLA consistently when network time allows it to at $\geq$100ms, while the greedy approaches fails to do so.
As such, \modiselect improves effective accuracy \emph{safely} as the SLA target increases.
Meanwhile, by selecting a model probabilistically \modiselect absorbs the network variation
\emph{better} by using a diverse set of CNN models.
}
\label{eval:fig:two-models}
\end{figure}

\subsubsection{Prototype evaluation}
\label{sec:eval:prototype}

In Figure~\ref{eval:subfig:prototype:trade-off}, we plot the percentage of inference requests that exceed the SLA (left y-axis) and the percentage of inference requests that are correctly classified (right y-axis) for different SLA targets.
We further annotate the figure with two important timelines: on-device inference time with \texttt{MobileNetV1 0.25} and on-device inference time with \texttt{MobileNetV1 1.0} (from left to right) to better illustrate \modiselect performance. The mobile device is connected to our campus WiFi which has an average network time of $63$ms over the course of the test.

As we can see, \modiselect is able to gradually reduce the percentage of SLA misses as the SLA target increases. In particular, we start to observe reduction in the number of SLA violations and improved inference accuracy once the SLA target is larger than $115$ ms. This is due to \modiselect recognizing that the time budget is extremely small and utilizing a low-latency model, \texttt{MobileNetV1 0.25}. As the SLA increases further, the overall inference accuracy begins to improve but still exhibits some variation. The improved accuracy is due to \modiselect identifying the increased time budget and beginning to use the more accurate \texttt{InceptionV3} model while the continuing variation in accuracy is due to \modiselect accounts for network variability and occasionally chooses \texttt{MobileNetV1 0.25}.

\emph{\textbf{Result:} \modiselect is able to adapt its model selection with the goal to minimize SLA violations while improving inference accuracy, even when SLA target is set to be as low as executing a mobile-optimized model on-device.}

\subsubsection{Benefits over a greedy model selection}
\label{sec:eval:benefits-of-multimodel-hosting}

To examine \modiselect's ability to handle the trade-offs between inference response time and accuracy, we compare \modiselect to a \emph{greedy} algorithm that always chooses the most accurate CNN model for a given SLA. In Figure~\ref{eval:subfig:two-models:compare}, we plot the average end-to-end inference time (left) and inference accuracy achieved by these two algorithms.
This figure shows that \modiselect consistently achieves up to 42\% lower inference latency, compared to \emph{greedy}. Moreover, \modiselect can operate under a much more stringent SLA target ($\mathtt{\sim}115ms$) while \emph{static greedy} continues to incur SLA violations until SLA target is more than 200ms. The key reason is because \modiselect is able to effectively trade-off accuracy and inference time by choosing from a diverse set of models (see Figure~\ref{eval:subfig:two-models:model-usage}). Consequently, \modiselect has an accuracy of 68\% (on par to using \texttt{MobileNetV1 0.75} which can take 2.9x more time running on mobile devices) under low SLA target ($\mathtt{\sim}115ms$), but is able to match accuracy achieved by \emph{greedy} when SLA target is higher. Note that even though \emph{static greedy} achieves up to 12\% higher accuracy, it does so by sacrificing inference latency.

In Figure~\ref{eval:subfig:two-models:model-usage}, we further analyze \modiselect performance by looking at its model usage patterns under different SLA targets. At very low SLA target ($<$ 30ms), \modiselect aggressively chooses the fastest model \texttt{MobileNetV1 0.25} since none of the managed models satisfy Equation~\eqref{eq:c1} and~\eqref{eq:c2}. As the SLA target increases, \modiselect explores more accurate but slower models than \texttt{MobileNetV1 0.25}. There are two key observations: (1) \modiselect is effective in picking the more appropriate model to increase accuracy while staying safely within SLA target. For example, \texttt{InceptionResNetV2} is never selected by \modiselect because better alternatives \texttt{InceptionV3} for lower SLA target and \texttt{InceptionV4} for higher SLA target exist. (2) \modiselect faithfully explores eligible models and is able to ``converge" to the most accurate model when SLA target is sufficiently large.

\emph{\textbf{Result:} \modiselect outperforms \emph{greedy} with up to 43\% end-to-end latency reduction, while is able to keep up with accuracy with SLA budget is larger than $250$ms. The key reason is because \modiselect is able to adapt its model selection by considering both the SLA target and network transfer time, while \emph{greedy} naively selects the most accurate model.}

\section{Related Work}
\label{sec:related}
To keep up with the increasing popularity of using deep learning within mobile applications~\cite{2019www:firstlook}, there has been a wide range of work on providing efficient mobile deep inference. These efforts range from optimizing mobile-specific models to improving the performance of inference serving systems.

\para{On-device execution.} Efforts for enabling executing deep learning models
directly on mobile devices fall in two broad categories: mobile-specific model
optimizations and redesigning mobile deep learning frameworks.

Concretely, researchers have investigated various ways to make DNN models
efficient~\cite{DBLP:journals/corr/SzeCYE17}. First, post-training optimizations
such as quantization uses simpler representations of weights and bins weights to
improve compressibility~\cite{tensorRT, DBLP:journals/corr/HanMD15} allowing for
reduced load time. Second, techniques such as
pruning~\cite{DBLP:journals/corr/HanMD15}, removing model weights with low
contributions, reducing the number of computations needed for inference as well
as model sizes.
Third, redesign of networks can also lead to improved inference time. An early example was the mobile-specific SqueezeNet~\cite{DBLP:journals/corr/IandolaMAHDK16} and this trend has continued with MobileNet~\cite{DBLP:journals/corr/HowardZCKWWAA17} which was designed as a compact alternative to the complex InceptionV3 model~\cite{DBLP:journals/corr/SzegedyVISW15}.

To enable running models across different hardware architectures~\cite{17emdl:mobirnn}, researchers
have redesigned deep learning frameworks~\cite{Lane:2016ko,deepsense:2016,2016mobisys:mcdnn} with the goal of providing optimized
runtimes. For instance TensorFlowLite~\cite{google_tensorflow:lite} and
Caffe2~\cite{caffe2} both leverage mobile-specific optimizations that allows
deep learning models to execute smoothly on mobile hardwares. Recently,
researchers investigated system-level optimizations for supporting multiple
on-device mobile applications~\cite{2018:nestdnn,2018mobicom:deepcache,19mobisys:deqa}. Our work can leverage these proposed optimizations to further improve mobile
deep inference performance by judiciously selecting models at runtime.

\para{Remote execution.} Cloud-based solutions have demonstrated their
effectiveness in handling traditional workloads~\cite{Guo2017mmsysJ}. Code
offloading~\cite{maui, olie} has been widely used to augment the performance of
mobile applications with constrained hardware resources. Due to the reliance on
network connectivity, code offloading is often done at runtime~\cite{maui}.
Determining the optimal partition of computation graphs can be solved
optimally~\cite{maui} with approaches such as Integer Linear Programming (ILP).
However, these optimal solutions fall short because they assume access to prior
performance information, such as execution time and energy~\cite{maui, olie} and
often incur long decision time. \modiselect leverages the key idea of runtime
computation offloading by selecting from different deep learning models, for
both inference speed and accuracy gain.

Recently, various model serving platforms~\cite{tensorflow_serving, clipper,
DBLP:journals/corr/ChenLLLWWXXZZ15} provides web-based services
that mobile applications can leverage. Further, researchers have started to
understanding the performance and cost trade-offs when running inference
services in the cloud~\cite{2019arxiv:clouddnn,2019atc:mark,2018ic2e:dnnserveless}. These platforms are often designed with
the key focus of managing model lifecycle from training to deployment, providing
low-latency and high-throughput serving systems, and cost-effective cloud
resource managements. These projects are beneficial to \modiselect as they provide
infrastructure supports for hosting a range of models. Moreover, \modiselect
complements these works with system and algorithm designs that gear towards mobile applications.

\section{Conclusion}
\label{sec:conclusion}

In this paper, we conducted comprehensive empirical measurements that geared towards understanding the performance implications of
running deep learning models on mobile devices and in the cloud. We identified a number of key performance factors, such as mobile networks and CNN models, and demonstrated the need of cloud-based inference, especially for complex CNN models and older mobile devices. Towards mitigating the impact of the mobile network variations on cloud-based inference performance, we proposed \modiselect
that manages a set of CNN models and uses probabilistic-based to adapt its model selection to the heterogeneous mobile requirements.
Our evaluations show that \modiselect is able to transparently switch between CNN models as SLA relaxes, and that \modiselect improves SLA attainment by 88.5\% while achieving comparable accuracy compared to greedy algorithms.

\balance
\bibliographystyle{IEEEtran}
\bibliography{modi_arxiv.bib}
\vspace{-10mm}
\begin{IEEEbiography}[{\includegraphics[width=1in,height=1.25in,clip,keepaspectratio]{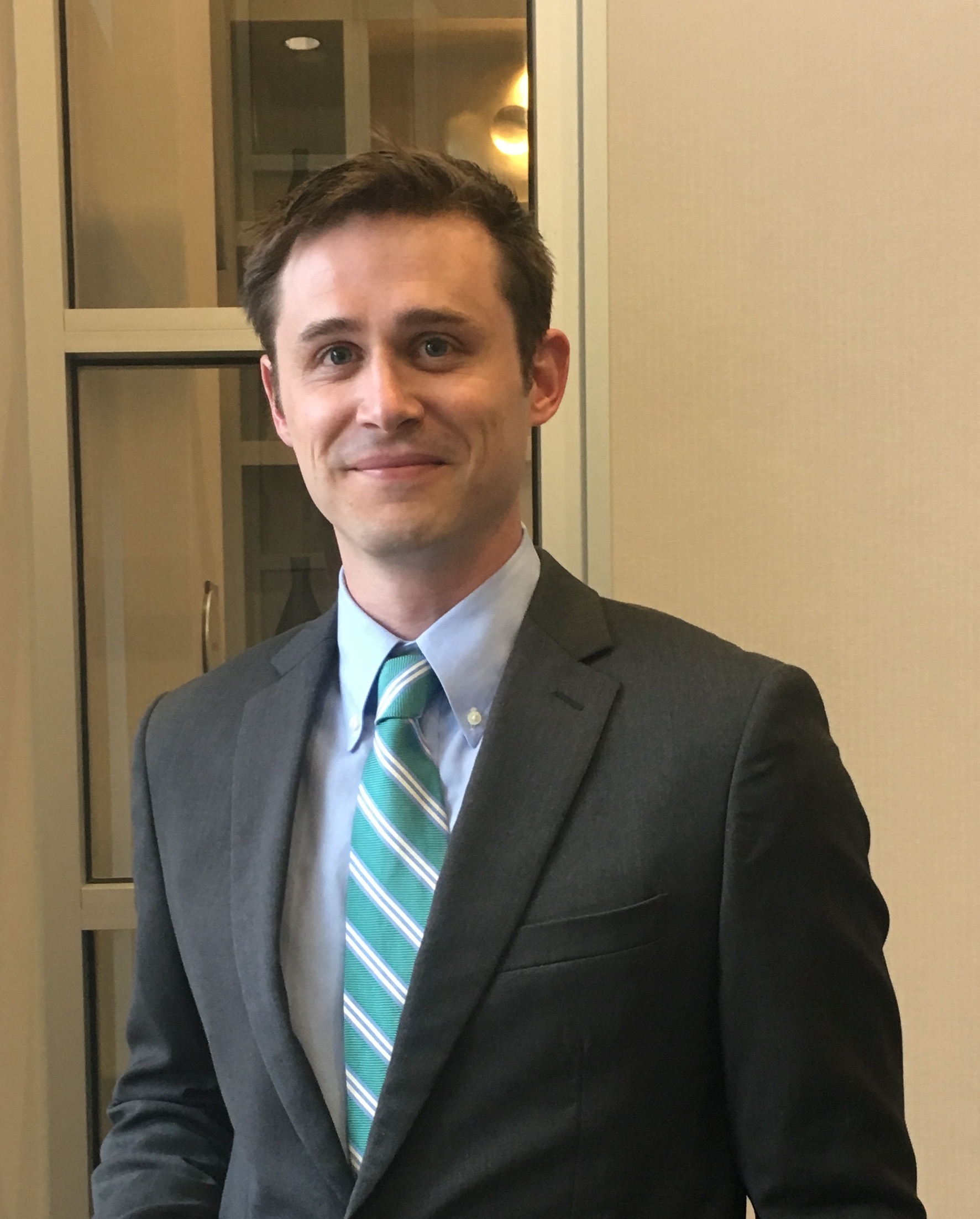}}]{Samuel S. Ogden}
is a Ph.D. student at Worcester Polytechnic Institute.
He received his M.S. in Computer Science from the University of Vermont in 2013 and his B.S in Pure Mathematics and Electrical Engineering in 2010.
His research interests include cloud and mobile computing and deep learning systems.
Contact him at ssogden@wpi.edu.
\end{IEEEbiography}

\begin{IEEEbiography}[{\includegraphics[width=1in,height=1.25in,clip,keepaspectratio]{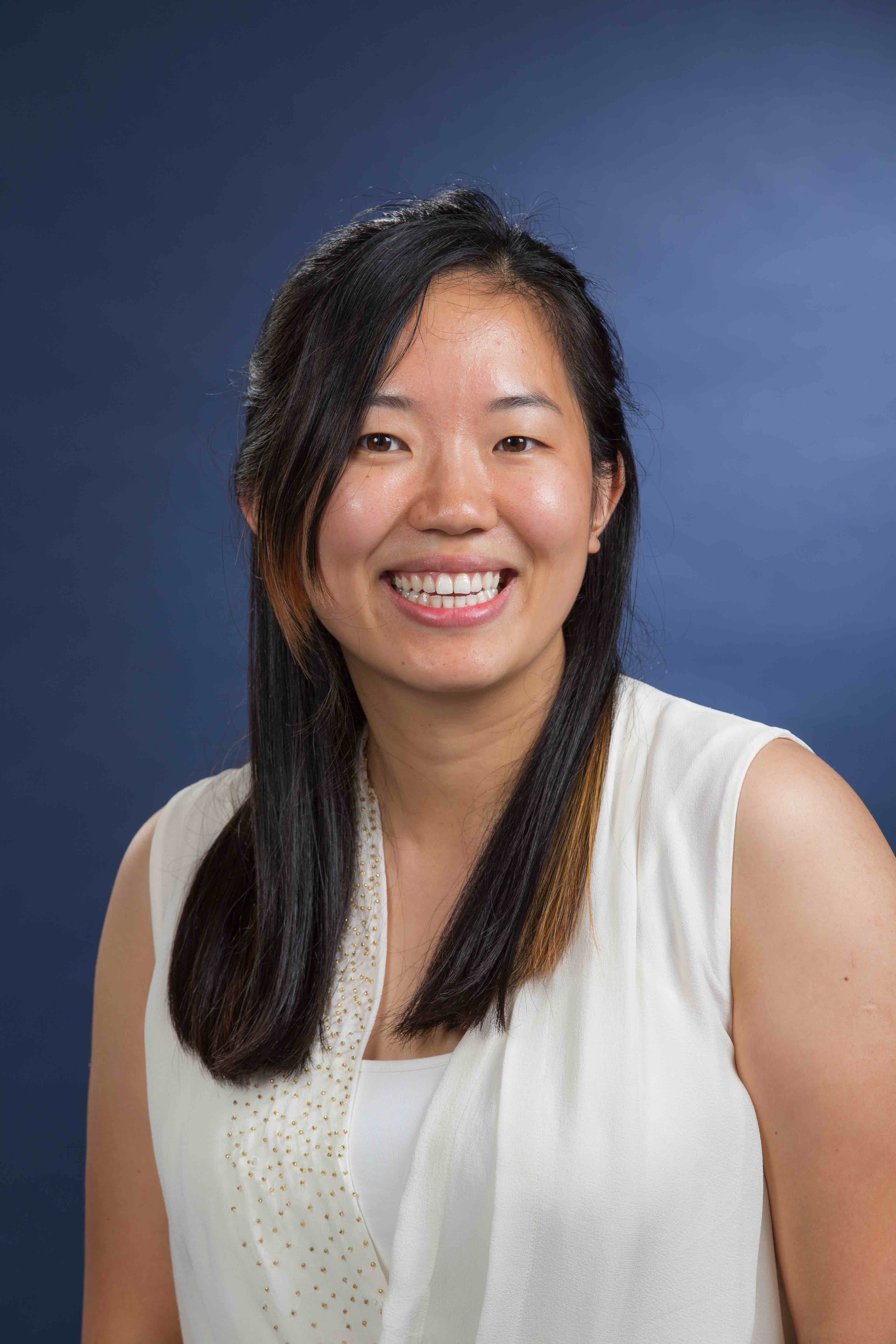}}]{Tian Guo}
is an Assistant Professor in
the Computer Science Department at Worcester
Polytechnic Institute. She received her Ph.D. and
M.S. in Computer Science from the University of
Massachusetts Amherst in 2013 and 2016, respectively, and her B.E. in Software
Engineering from Nanjing University in 2010. Her research
interests include distributed systems, cloud computing and mobile computing. Contact her at
tian@wpi.edu.
\end{IEEEbiography}
\vfill

\end{document}